\documentclass[]{aastex631}

\graphicspath{{./}}
\usepackage{wrapfig}
\usepackage{enumitem}
\usepackage{upgreek}
\begin{document}

\title{A Catalog of the Highest-Energy Cosmic Rays Recorded During Phase I of Operation of the Pierre Auger Observatory}

\email{spokespersons@auger.org}

\collaboration{0}{The Pierre Auger Collaboration}
% created on 2022-07-26

\begin{abstract}

A catalog containing details of the highest-energy cosmic rays recorded through the detection
of extensive air-showers at the Pierre Auger Observatory is presented with the aim of opening
the data to detailed examination. Descriptions of the 100 showers created by the highest-energy
particles recorded between 1 January 2004 and 31 December 2020 are given for cosmic rays that
have energies in the range 78\,EeV to 166\,EeV. Details are also given of a further nine very-energetic events that have been used in the calibration procedure adopted to determine the
energy of each primary. A sky plot of the arrival directions of the most energetic particles is
shown. No interpretations of the data are offered.

\end{abstract}


%% and \citet commands to identify citations. The citations are

\section{Introduction} \label{sec:intro}

The energy spectrum of cosmic rays extends to beyond 100\,EeV. Where and how these particles, dominantly the nuclei of the common elements up to iron, are accelerated is one of the major puzzles of astroparticle physics.
The flux above 50\,EeV is about 0.5 particles per km$\mathrm{^{2}}$ per century, so that measuring their properties requires the detection of the cascades or \textit{air showers} that the particles create in the atmosphere.
In this paper, the methods used by the Pierre Auger Collaboration to obtain the arrival directions and energies of the 100 highest-energy particles in the range 78\,EeV to 166\,EeV are outlined, and details of the main features of the air showers produced by the cosmic rays are presented.
Phase I of operation of the Observatory ended on 31 December 2020. It is thus timely to release a catalog to demonstrate the quality of the data that lie behind measurements of the energy spectrum, the distribution of arrival directions, and the mass of the highest-energy cosmic rays that have been reported elsewhere (\cite{PierreAuger:2020qqz}, \cite{PierreAuger:2017pzq}, \cite{PierreAuger:2014sui} and \cite{PierreAuger:2014gko}). 
The events discussed here are included in the data set recently used in a discussion of the arrival directions of events above 32\,eV (\cite{PierreAuger:2022axr})\footnote{Two events with energies close to 100\,EeV, used in a recent study of mass composition (\cite{Yushkov:2020nhr}), are not included here, or in \cite{PierreAuger:2022axr}, as different selection criteria were adopted.}.
No interpretations of the data are offered in this paper.  Recent reviews, together with  some interpretations, of data on high-energy cosmic-rays can be found in \cite{Mollerach} and in \cite{Batista}.  A discussion of present data on the highest-energy cosmic-rays is included in the US Community Study on the Future of Particle Physics 2021 (\cite{Coleman}).
\\

\noindent
The structure of the paper is as follows. In Section~\ref{sec:detection}, after a brief outline of the methods used to detect the highest-energy cosmic rays, the instrumentation of the Auger Observatory, relevant to this paper, is described. In Section~\ref{sec:reconstruction}, brief accounts of the techniques developed by the Collaboration are given, including that used to assign the energy of the primary particle that initiates each air shower, or event. In Section~\ref{sec:catalog}, the catalog is described and some events within it are discussed in detail. These descriptions have been prepared to aid scrutiny of the complete sample publicly available at \url{https://opendata.auger.org/catalog/}. In Section~\ref{sec:skyMap}, a sky map of the arrival directions of the events is shown.

\clearpage

\section{The detection of high-energy cosmic rays and the Pierre Auger Observatory} \label{sec:detection}

\subsection{The Detection of High-Energy Cosmic Rays}
Above an energy of about 100\,TeV, the flux of cosmic rays is so low that detectors flown using balloons, or deployed in space, are insufficiently large to detect useful numbers of primary particles directly. At higher energies, the particles create cascades, largely of electrons, positrons, photons, and muons, that propagate through the atmosphere as extensive air-showers. Such showers can be detected through the charged particles and photons that reach ground level, and by observing light emitted from the atmosphere. Properties of the primary cosmic rays are inferred from studies of these showers.\\

\noindent
If the incoming particle is a proton or an atomic nucleus, then, in the first interaction with a nucleus in the atmosphere (usually nitrogen or oxygen), hundreds of pions are created. The neutral pions decay rapidly into photons that initiate electromagnetic cascades through pair production, with the electrons and positrons subsequently producing bremsstrahlung radiation. The electromagnetic cascade grows until the rates of energy loss through these two processes are exceeded by the rate of energy loss by ionization. Charged pions interact with nuclei to produce additional pions that further enrich the cascade until their energy falls below $\sim$300\,GeV when charged-pion decay becomes more probable than interaction with nuclei. The nucleons of the incoming primary lose, on average, about 50\% of their energy in the first interaction, and in further similar interactions, thus enhancing the number of secondary particles in the shower. The charged particles and the accompanying photons spread out laterally because of scattering, and because of the transverse momentum of the particles produced in the collisions. \\

\noindent
The shower of secondary particles can be detected in several ways. One method is to spread a number of detectors over the ground: currently scintillation counters or water-Cherenkov detectors are the most widely adopted. At ${\sim}1$\,PeV, the footprint of the shower is about 10$\mathrm{^{4}}$\,m$\mathrm{^{2}}$, while, for the energies of interest here, the equivalent scale is many square kilometres. The number of detectors deployed in any shower array is, of necessity, a compromise dictated by cost.\\

\noindent
The particles of the shower can be thought of as traveling at close to the speed of light in a slightly curved, disc-like, configuration similar to a giant dinner-plate, with the density of particles falling-off rapidly from the centre of the disc. The fall-off is described by a \textit{lateral distribution function} (LDF), knowledge of which is important for the reconstruction of events. 
The zenith angle of a shower is determined typically to $\sim$1$^{\circ}$ from the times of arrival of the first particles in the shower-disc at the detectors. \\

\noindent
Other methods of shower detection make use of the fluorescence radiation that results from the excitation of molecules of atmospheric nitrogen by the charged particles in the shower and of the Cherenkov light created as these particles cross the atmosphere. Fluorescence radiation is emitted isotropically and can be observed at large distances from the shower. Detection is technically demanding as only about 5.6~photons are emitted in the 300 to 400\,nm band for each MeV of energy deposited (\cite{AIRFLY:2012msg}). The challenge of detecting such light from a shower produced by a particle of $\sim$3\,EeV at~15\,km is akin to trying to observe a 5-Watt light bulb moving at the speed of light at this distance. By contrast, Cherenkov radiation is much brighter with around 30~photons emitted between 400~and~700\,nm per metre of track (\cite{Galbraith}), with the light concentrated along the direction of travel of the shower and with a lateral spread dictated by that of the electrons. In the events described below, scattered Cherenkov light is a background that must be accounted for when reconstructing the properties of showers with the fluorescence detectors.\\

\noindent
Other aspects of shower detection, specific to the Auger Observatory, are discussed in Section~\ref{sec:auger}.

\subsection{The Pierre Auger Observatory} \label{sec:auger}
The Pierre Auger Observatory is the largest cosmic-ray detector ever constructed. It was designed to explore the properties of the highest-energy cosmic rays with unprecedented statistical precision and this objective has been achieved. The primary experimental targets were the determination of the energy spectrum, the distribution of arrival directions and the mass composition of cosmic rays above $\sim$1\,EeV. Studies of lower-energy cosmic rays, of particle physics, and of geophysical phenomena, now form important additions to the scope of the project. \\

\begin{figure}[t!]
 \epsscale{0.7}
\plotone{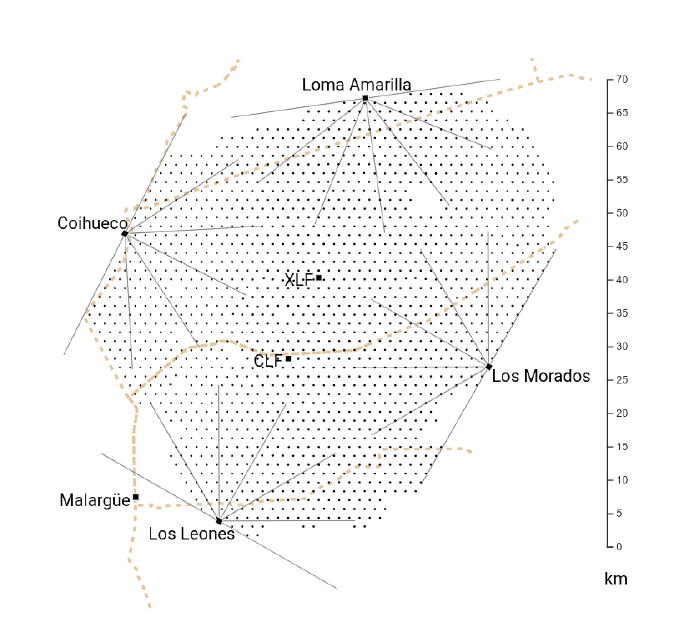}
\caption{The layout of the Pierre Auger Observatory covering 3000\,km$\mathrm{ ^{2}}$. Each small dot corresponds to a water-Cherenkov detector. Fluorescence detectors are located at Los Leones (LL), Coihueco (CO), Loma Amarilla (LA) and Los Morados (LM). The 30$^{\circ}$ azimuthal fields of view of the six telescopes at each site are shown by the radial lines emanating from them: the vertical reach of the telescopes extends to an elevation of 28.6$^{\circ}$. Data are transmitted to the central laboratory, located at a campus in Malargüe, using a purpose-built communication network. The dashed lines show roads. Gaps in the layout of the array arise due to difficulties with landowners. Steerable lasers (see text) are located at the positions CLF and XLF.}
\label{fig:mapAuger}
\end{figure}

\noindent
The Observatory is located near the city of Malargüe, Mendoza Province, Argentina, between latitudes 35.0$^{\circ}$S and 35.3$^{\circ}$S and longitudes 69.0$^{\circ}$W and 69.4$^{\circ}$W. The mean altitude of the site is about 1400\,m above sea-level, corresponding to an atmospheric overburden of about $875\,\text{g}/\text{cm} ^{2}$. The Observatory comprises an installation of about 1600 water-Cherenkov detectors, separated by 1500\,m, laid out on a triangular grid over an area of 3000\,km$\mathrm{ ^{2}}$ (the Surface Detector, SD), and overlooked by a Fluorescence Detector (FD) comprising four stations, each containing 6 telescopes, each with 440 photomultipliers and a 13\,m$\mathrm{ ^{2}}$ mirror. A map of the site, showing the features relevant to this paper, is presented in Figure~\ref{fig:mapAuger}. A detailed description of the instrumentation can be found in \cite{PierreAuger:2015eyc}.\\

\noindent
The water-Cherenkov detectors (each $10\,\text{m}^{2}\times~1.2\,\text{m}$) are used to measure the energy flow at the ground level carried by the flux of muons, electrons, positrons, and photons in the air showers generated by the primary particles. In near-vertical events, there are 10 times as many photons as electrons and positrons, which in turn exceed the number of muons by about the same factor. The average energy of the muons in a near-vertical shower is $\sim$1 GeV, while the mean energy of the entities of the electromagnetic component is $\sim$10\,MeV. Thus, the electromagnetic radiation is largely absorbed in the 3.2 radiation lengths of the 1.2\,m depth of the water-Cherenkov detectors, whereas most of the muons pass straight through, losing energy only through ionization. The energy deposited in the water by the shower components is expressed in terms of the signal, measured using three 9 inch photomultipliers, from a muon traversing vertically and is expressed in terms of ‘Vertical Equivalent Muons’, or VEM, and corresponds to an energy deposit of $\sim$250\,MeV. In a vertical shower produced by a particle of 10\,EeV, the signal, \textit{S}(1000), at 1000\,m from the densest region of the shower, called the \textit{core}, is $\sim$40\,VEM, and is roughly a 50/50 mixture of signals from muons and the electromagnetic component.\\

\noindent
The times of arrival of particles at the water-Cherenkov detectors are measured using GPS signals that are also exploited to locate the position of each detector to 20\,cm and 50\,cm in the horizontal and vertical directions, respectively. At the highest energies, the incoming direction can be determined to better than 0.4$^{\circ}$ (\cite{PierreAuger:2020yab}).\\

\noindent
The thickness of the shower disc (in nanoseconds) is defined as the time that it takes for the signal amplitude to grow from 10 to 50\% (this time, t$\mathrm{_{1/2}}$, is referred to as the \textit{risetime}). In events that arrive nearly vertically, risetimes vary from a few nanoseconds close to the core, to $\sim$300\,ns at distances of $\sim$1\,km, and decrease as the zenith angle increases.\\ 

\noindent
The time-profiles of the signals recorded with the water-Cherenkov detectors have been used in several studies. It has been possible to build observables that allow inferences to be made about the mass composition, and to probe hadronic interactions above the energies attained at the Large Hadron Collider, with a statistical sample of $\sim$81,000 events (\cite{PierreAuger:2017tlx}). Additionally, searches for photons and neutrinos in the cosmic particle flux have been made (\cite{PierreAuger:2016ppv}, \cite{PierreAuger:2019ens}). Above 60$^{\circ}$, the risetimes of the signals are too fast to be measured accurately with the electronics currently in use.\\

\noindent
Measurements of the fluorescence light make possible a calorimetric estimate of the energy of the primary particle (\cite{PierreAuger:2020qqz}) and provide a key tool used in the determination of the mass of the primary particles (\cite{PierreAuger:2014gko}). For such studies, it is essential to monitor the atmosphere and this is done using steerable lasers located at the positions marked CLF and XLF in Figure~\ref{fig:mapAuger} (\cite{PierreAuger:2012rhm}). These lasers are also used to make independent checks on the accuracy of the reconstruction of the arrival directions possible (\cite{Mostafa}).\\

\noindent
Data taking began on 1~January~2004 with 154 water-Cherenkov detectors and two fluorescence stations partly operational. Observations with the instrumentation of Figure~\ref{fig:mapAuger} started in June 2008 and have been in progress ever since. The surface detector is operated almost continuously, while observations with the fluorescence detector are restricted to clear dark nights. Phase I of the project was completed on 31 December 2020. Instrumentation used in other Phase I studies are described in are described in \cite{PierreAuger:2015eyc}. It is thus timely to release a catalog giving details of the extensive air-showers produced by the highest-energy cosmic rays observed thus far. In addition to the detailed information on the 100 events of the highest energy recorded between 1~January~2004 and 31~December~2020, which are part of the set of events discussed by \cite{PierreAuger:2022axr}, nine events of slightly lower energy, used for energy calibration, have been included to increase the number of fluorescence events presented. \\

\section{Reconstruction of shower parameters} \label{sec:reconstruction}

The properties that can be determined most directly are the arrival direction and the energy of the primary particle that initiates each air shower. Estimating the mass of the incoming particle is more complex as it requires assumptions to be made about the hadronic physics associated with interactions of nucleons and pions and, at present, it is not possible to identify the mass of the primaries except on an average basis (e.g., \cite{PierreAuger:2014gko}). No discussion of measurements relating to mass determination is included in this paper. In the following sections, brief descriptions of the methods used to find the arrival directions and the energies are given.

\subsection{Recording of the data} \label{sec:recording}

Data from the surface detectors to be used in reconstruction are derived from a relatively complex triggering procedure described in \cite{PierreAuger:2010zof}. Briefly, triggers from each station, tagged with the GPS time, are sent at a rate of $\sim$20\,Hz to a computer located at the campus in Malargüe (Figure~\ref{fig:mapAuger}) via a purpose-built link for communications. The computer is used to search for spatial and temporal coincidences between triggers from the detectors. When a coincidence is found between at least three stations, data from triggered detectors are downloaded (\cite{PierreAuger:2010zof}). In addition to the trigger-time, the data include read-outs from Flash Analog-to-Digital converters (FADCs) associated with each of the three photomultipliers in the water-Cherenkov detectors. GPS time stamps have a precision of 12\,ns, while the FADCs are 10-bit running at 40\,MHz. From the FADC information, the amplitude and time structure of each signal are obtained.\\

\noindent
Data from the fluorescence detectors are recorded in a different manner (\cite{PierreAuger:2009esk}). The telescopes at each of the four fluorescence stations are operated remotely from the Malargüe Campus or, since 2017, additionally from various locations around the world. The camera of each telescope contains 440 photomultipliers (pixels): the recording of signals and time-stamps is completely independent of that used for the surface detectors. A very loose criterion of a localized pattern of four pulses in consecutive time order is adopted as the trigger at each fluorescence telescope. Those triggers where a shower track can be found are transmitted to the central computer, together with information on the geometry of the shower candidate. From this information, the time of impact of the shower at a ground position in the region of the surface detectors is computed, so that all FADC traces in the region, arriving within 20\,${\upmu}$s, are also centrally recorded. After each night of operation, data from the fluorescence triggers are then merged with those data collected with the surface detectors: these form the \textit{hybrid} data set. For high-level analyses, several quality cuts are applied to the fluorescence events, including those relating to cloud cover and atmospheric aerosols. Further cuts are made to ensure that the selection of events is unbiased with respect to primary particle mass (\cite{PierreAuger:2014sui}). The overall efficiency of these cuts is such that approximately 25\% of SD events with energies above 10\,EeV, registered during FD operation, have an accompanying good quality and unbiased FD shower profile.

\subsection{Reconstruction of the arrival direction and energy of showers} \label{sec:arrival}
\subsubsection{Introduction} \label{sec:recoIntro}

While the reconstruction of the arrival direction of an air shower is relatively straight-forward, as outlined in Section~\ref{sec:verticalReco}, the determination of the parameter of the shower adopted as a surrogate for primary energy is more difficult. This is because, as the zenith angle increases, the shower loses the near-perfect circular symmetry found in an event generated by a cosmic ray entering the atmosphere at 0$^{\circ}$. The loss of symmetry of the distribution of the signal size in the plane perpendicular to the arrival direction of a shower arises for several reasons: from geometrical effects associated with the angles at which high-energy particles are emitted in early interactions, from geometrical effects relating to the direction of travel of particles entering the detectors, from attenuation - particularly of the electromagnetic component - as the shower crosses the array, and from the effect of the geomagnetic field. The most direct experimental evidence of asymmetry is found in studies of the risetime of the signals from the water-Cherenkov detectors (\cite{PierreAuger:2016tar}).\\

\noindent
The consequences of asymmetries of the signal sizes have been studied in some detail using simulations. \cite{Luce:2021ode} have examined the impact on the electromagnetic component. At 1000\,m from the shower axis, the amplitude of the asymmetry of the signal size is $\sim$50\% in a shower produced by a primary of 10\,EeV at a zenith angle of 45$^{\circ}$. However, estimates of the parameter used to define the shower size (the signal size at 1000\,m from the shower axis, \textit{S}(1000) - see below) are changed by less than 10\%. This is largely because the contribution of muons to the total signal in a detector rises with increasing zenith angle. \\

\noindent

\begin{figure}[ht!]
% \epsscale{0.4}
\centering
\includegraphics[trim={2cm 2.0cm 0 2cm},width=1\textwidth]{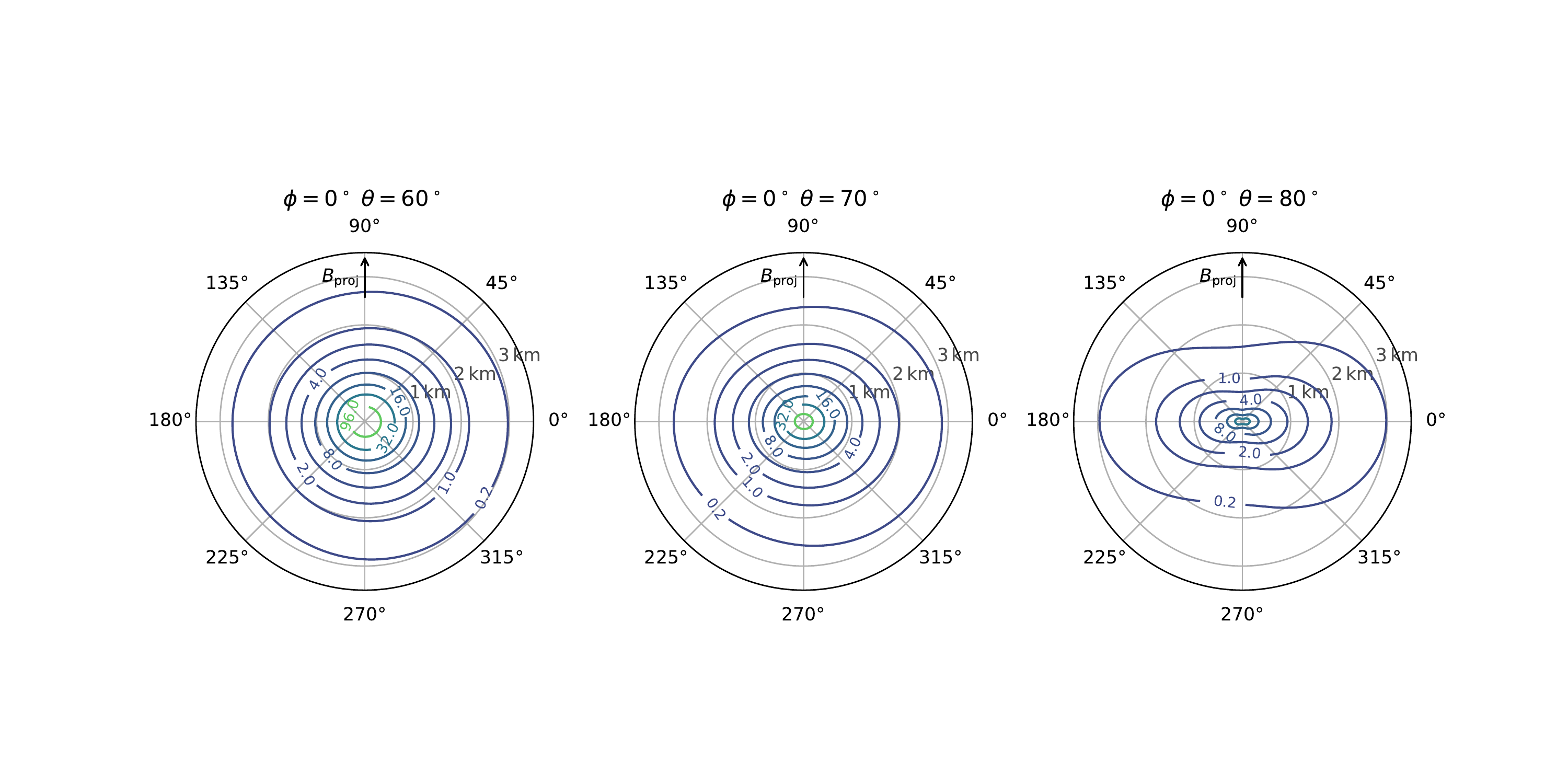}
\caption{Parameterized densities of muons for a 10\,EeV proton shower at zenith angles of 60$^{\circ}$, 70$^{\circ}$ and 80$^{\circ}$ arriving
from azimuth, $\phi$ = 0$^{\circ}$. Radial units are in kilometers. The coordinate system is defined in the plane perpendicular to
the shower direction with the y-axis parallel to the projection of the Earth’s magnetic field, $B_{\rm{proj}}$, on that plane. The
magnitudes of the muon densities are indicated (32, 16, 8... per m${^2}$).}
\label{fig:muonMap}
\end{figure}

At relatively small zenith angles, simulation studies have also been used to show that the effect of the geomagnetic field changes estimates of \textit{S}(1000) by only a few percent for angles around 45$^{\circ}$ (\cite{PierreAuger:2011yxe}). However, as the zenith angle increases, the effect of this field becomes more evident because of the increasingly long path-length of the muons as they cross the atmosphere. In Figure~\ref{fig:muonMap} the densities of muons reaching the ground, again estimated through simulation, are shown for three zenith angles.\\

\noindent
It is evident that the asymmetry of the radial distribution of the muons in the shower increases with zenith angle, becoming particularly apparent above 70$^{\circ}$. At such angles, the electromagnetic part of the shower, arising dominantly from the decay of neutral pions, has been largely absorbed as the atmospheric thickness exceeds $2440\,\text{g}/\text{cm} ^{2}$. However, an electromagnetic component, arising from muon-bremsstrahlung, knock-on processes and muon decay, is present and is time-synchronous with the muons, so that the time-spread of the signals is small, as will be seen in events discussed in Section~\ref{sec:catalog}.\\

\noindent
Novel methods have been developed to analyze events of large zenith angle (\cite{Ave:2000nd}, \cite{PierreAuger:2014jss}) as discussed in Section~\ref{sec:inclinedReco}. There is, of course, no sharp transition between the
zenith angle range in which atmospheric absorption dominates and that in which geomagnetic
effects assume the greater importance. Above $\sim$60$^{\circ}$ the accuracy of reconstruction of both the
direction and energy are increasingly improved using the new techniques (\cite{Schmidt}), and
accordingly the different approaches have been adopted above and below this zenith angle.

\subsubsection{Reconstruction of events with zenith angle $<$ 60$^{\circ}$} \label{sec:verticalReco}

The methods used to reconstruct events with zenith angle, $\theta <$ 60$^{\circ}$ recorded by the water-Cherenkov detectors are described in detail by \cite{PierreAuger:2020yab}. The zenith angle is measured from the zenith while the azimuth angle, $\phi$, is measured counter-clockwise from East. For showers as large as those described here, all arrival directions are determined to better than 0.4$^{\circ}$. Accordingly, as deflections by the Galactic magnetic field of protons exceed this number, even for the energies discussed here, no uncertainties are given. An uncertainty of 0.4$^{\circ}$ in the zenith angle leads to an uncertainty in energy of $<$ 0.2\%.\\

\noindent
The positions of the detectors with respect to the core of the shower are found by fitting the observed signals to a lateral distribution function\footnote{
When the core of a shower falls close to a detector, the signal can be so large that the electronic recording channels may saturate. This usually occurs for detectors within about 500\,m of the core where the signal is greater than 1000\,VEM. An algorithm is used to estimate the true magnitude of the signal from the amplitude of the undershoot which is introduced capacitatively. Moreover, for signals larger than 2000\,VEM, the PMT response is highly non-linear so that only timing information is used and the signal is treated in the LDF fit only as a lower limit to the actual size of the signal. Note that the estimated true signal value is used in the LDF fit for saturated signals smaller than 2000\,VEM. For 50\% of the events contained in the full data set, the signal in one station is saturated: 3 of the events discussed below have two saturated stations. Examples of saturated signals can be found in Section~\ref{sec:catalog}, and in the larger data base.}. In general, because of the wide spacing of the detectors, it is not possible to determine this function for every event. Accordingly, an empirical description, based on the pioneering studies of \cite{Greisen1956}, \cite{Greisen1960} and \cite{Kamata}, has been adopted:

\[S_{\text{LDF}}(r) = \textit{S}(1000) \Biggl[\Biggl(\frac{r}{r_{\rm opt}}\Biggr) \Biggl(\frac{r+r_{\rm s}}{r_{\rm opt}+r_{\rm s}}\Biggr)\Biggr]^\beta\]
\noindent
with $r_{\rm s}$ fixed at 700 m. The slope factor, $\beta$, is negative, changing from about $-2.6$ at $\theta$ = 0$^{\circ}$ to about $-1.9$ at 60$^{\circ}$. The flattening of the lateral distribution function with increasing angle is largely due to the increasing dominance of the muon component.\\

\noindent
The quantity $r_{\rm opt}$ relates to the spacing of the detectors and is the distance at which uncertainties in the reconstructed signal size, arising from lack of knowledge of the lateral distribution function, is minimized (\cite{hillas1970}, \cite{hillas1971}). For the detectors of the Auger Observatory, where the spacing is 1500\,m, $r_{\rm opt}$ has been shown to be close to 1000\,m (\cite{Newton:2006wy}). The signal size at this distance, \textit{S}(1000), is used to estimate the primary energy. \\

\noindent
The average statistical uncertainty in the determination of \textit{S}(1000) at the highest energies is 8\% (\cite{PierreAuger:2020yab}). The uncertainty on the impact point is $\sim$50\,m. \textit{S}(1000) is influenced by changes in atmospheric conditions that affect the development of showers (\cite{PierreAuger:2017vtr}), and by the geomagnetic field that impacts on the signal sizes in the shower (\cite{PierreAuger:2011yxe}). Therefore, before using the shower-size estimator in the calibration procedure (Section~\ref{sec:energyReco}), corrections of $\sim$2\% and $\sim$1\% are made for the atmospheric and geomagnetic effects, respectively. 

\subsubsection{Reconstruction of events with zenith angles $>$ 60$^{\circ}$} \label{sec:inclinedReco}

The analysis of events with zenith angles $>$ 60$^{\circ}$ is important as extending measurements to these angles enhances the exposure of the Observatory by 30\%, and extends sky coverage to regions that would otherwise be inaccessible. However, as explained above, techniques different to those used to reconstruct showers arriving at smaller zenith angles must be adopted. 
Showers with zenith angles estimated to be as great as $\sim$90$^{\circ}$ have been recorded but, because the distance between detectors, as seen by the shower, is substantially shortened, the accuracy of reconstruction of the direction is badly degraded, and we restrict selection to those with $ \theta <$~80$^{\circ}$, where the directional uncertainties are~$<$~1$^{\circ}$. The procedures developed to analyze these events are discussed in detail in \cite{PierreAuger:2014jss}.\\

\noindent
Above 70$^{\circ}$ most of the particles at detector level are energetic muons accompanied by an electromagnetic component in equilibrium with the muons arising through bremsstrahlung, knock-on electrons and muon decay processes, which makes up 25\% of the signal beyond $\sim$1\,km from the core and around 30\% within 1\,km. Except at extreme distances, approximately 80\% of the signal arrives within about 200\,ns (see Figures~\ref{fig:PAO200313}~and~\ref{fig:PAO140131_1} in Section~\ref{sec:catalog} below). The muons travel tens to hundreds of kilometers before detection and are deflected significantly by the geomagnetic field. Thus, at ground level, the near-cylindrical symmetry associated with near-vertical events is lost, as shown in Figure~\ref{fig:muonMap}.\\

\noindent
For showers with an inclination between 60$^{\circ}$ to 70$^{\circ}$, and in particular at distances closer than 1\,km to the shower core, there is still a significant contribution from the electromagnetic component, 67\% at 60$^{\circ}$ and 100\,m, and accordingly this is included in the reconstruction (\cite{VALINO2010304}).\\

\noindent
The number of stations satisfying the trigger conditions above 60$^{\circ}$ increases with $\sec\theta$ so that at 30\,EeV the average number is $\sim$25 at 60$^{\circ}$, while at 80$^{\circ}$ it is $\sim$45. The method used for reconstruction is based on fitting the signal pattern recorded to what is predicted from modeling the shower development.
The muon signal scales with energy as $\rho_{\mu}\,(r) \propto E^\alpha$ with $\alpha$ in
the range 0.90 to 0.95. The expected density of muons at the ground is given by
$\rho_{\mu}(r) =N_{19}\,\rho_{\mu,19}(r,\theta,\phi) $, where $N_{19}$ is, chosen by convention, as a measure of shower size using a reference shower model and comparing the signals to those expected from simulated showers of 10\,EeV with the same arrival direction. Simulations have shown that $\rho_{\mu,19}(r,\theta,\phi)$, at fixed zenith and azimuth angle, varies by only about 5\% for changes in the energy and mass of the primary particle (\cite{Dembinski}).\\

\noindent
The absolute value of $N_{19}$ depends on the choice of mass composition and hadronic model used in the simulation for the reference model, but the dependence is constant with energy and between the primaries (\cite{PierreAuger:2015xho}). This uncertainty does not impact the estimate of the primary energy because the constant shift is absorbed by the method used to determine the energy scale, as outlined in Section~\ref{sec:energyReco}.

\subsubsection{Reconstruction of events recorded with the Fluorescence Detectors} \label{sec:FDreco}

The Fluorescence Detectors provide calibration information from which the energies of the more abundant events obtained with the water-Cherenkov detectors alone can be derived. Measurements of the fluorescence emission also give details of the longitudinal development of air showers in the atmosphere, with the determination of the depth at which the deposition of energy is greatest, the shower maximum. This is a key measurement for mass estimation. Details of the reconstruction methods are discussed in \cite{PierreAuger:2009esk} and \cite{PierreAuger:2014sui} with only a brief description given here.\\

\noindent
The 440 pixels in each camera, illuminated by light from the air shower, are used to reconstruct a plane that includes the axis of the shower and the location of the telescope. Within this plane, a three-dimensional reconstruction of the arrival direction is obtained by determining the geometry from the arrival times of the shower light at each pixel, and from the time of the arrival of the shower particles at the water-Cherenkov detector closest to the core of the shower. This hybrid technique, implemented for the first time at the Auger Observatory, enhances the precision with which the shower geometry is determined: the direction is known to $\sim$0.6$^{\circ}$
(\cite{Bonifazi:2009ma}). The signal from each pixel is recorded in 100\,ns intervals and the time and amplitude data are used to delineate the profile of the shower development using techniques described by \cite{Unger:2008uq}. This method allows differentiation between the various sources of detected light, namely the fluorescence light, direct Cherenkov light, and light scattered from the Cherenkov beam into the fluorescence telescope from air molecules and aerosols. \\

\noindent
For each 100\,ns interval, the energy deposited in the slant-depth interval corresponding to the measured light flux is estimated. These individual estimates are fitted using the universal shower profile function described in \cite{Andringa:2011zz},

\[ f(X) = \Biggl(\frac{{\rm{d}}E}{{\rm{d}}X}\Biggr)_{\rm{max}} \Biggl( 1 + \frac{R}{L} \bigl(X - X_{\rm{max}}\bigr) \Biggr)^{1/R^2} \exp \Biggl(-\frac{X-X_{\rm{max}}}{RL}\Biggr)
,\]
\noindent
where $f(X)$ is the energy deposit in the slant-depth $X$ and $({\rm{d}}E/{\rm{d}}X)_{\rm{max}}$ is the energy deposit at shower maximum. $X_{\rm{max}}$ is the slant-depth of the maximum of the energy deposit, while \textit{R} and \textit{L} are shape parameters loosely constrained in the fit to the average of measured values (\cite{Dawson:2020bkp}). The universal shower profile function is a re-casting of the Gaisser-Hillas functional form (\cite{gaisser1977}): its adoption diminishes correlations between shape parameters.\\

\noindent

The energy of each event ($\mathrm{E_{FD}}$) is determined by integration under the area defined by the longitudinal profile, $f(X)$, that defines the rise and fall of deposition of energy by the shower in the atmosphere, with the addition of 20\% at 0.1\,EeV and 12\% at 100\,EeV respectively. This augmentation accounts for energy that is not deposited in the atmosphere but is carried into the ground largely by muons and neutrinos. The model-independent methods of determining this factor are discussed in \cite{PierreAuger:2019dhr}. Above 10\,EeV, the energy is determined with a statistical precision of 8\% and with a systematic uncertainty of $\sim$14\% (\cite{Dawson:2020bkp}).

\subsection{Determination of the energy of the primary particles} \label{sec:energyReco}
The methods by which data from the surface detectors are calibrated to obtain the energies of the primaries are detailed in \cite{PierreAuger:2020qqz}. Use is made of hybrid events, both for showers with $\theta$ $<$ 60$^{\circ}$ (referred to as ‘vertical events’) and for events of larger zenith angles (‘inclined events’). \\

\noindent
For the vertical events, the measure of \textit{S}(1000) is first adjusted to the value that a shower would have had, had it arrived at 38$^{\circ}$ from the vertical, $S\mathrm{_{38}}$, as this is the median zenith angle for the vertical sample. Using the 3338 hybrid events that are available, the calibration relationship is $E\mathrm{_{FD}} =A\,{S\rm{_{38}}}^B$, with $A = (0.186 \pm 0.003)\,\text{EeV}$ and $B$~=~1.031~$\pm$~0.004. The calibration constants $A$ and $B$ are then used to estimate the energy for all SD events, $E\mathrm{_{SD}}$. The statistical uncertainty of $E\mathrm{_{SD}}$, obtained by propagating the errors on $A$ and $B$, is 1\% at the energies considered in this paper. The energy resolution, obtained from the spread of $E\mathrm{_{SD}}$ values at a given $E\mathrm{_{FD}}$ in the calibration events, is $\sim$8\% at the highest energies (\cite{PierreAuger:2020qqz}).\\

\noindent
A similar calibration procedure is adopted for the events with $\theta >$ 60$^{\circ}$. Here the calibration is made using $N\mathrm{_{19}}$ as the surrogate for the shower size. The value of $N\mathrm{_{19}}$ is then adjusted to the value ($N\mathrm{_{19,68}}$) for a shower arriving with 68$^{\circ}$, the median zenith angle of the sample. The calibration is made with 389 events and the values of $A$ and $B$ are $A$ = (5.32 $\pm$ 0.07)\,EeV and $B$~=~1.05~$\pm$~0.02, where $N\mathrm{_{19}}$ replaces $S\mathrm{_{38}}$. The smaller number of events available for evaluation of the energy of the more inclined events arises from the higher energy threshold required (4\,EeV as against 3\,EeV), and because there is a requirement for the shower maximum to be in the field of view of the FD telescopes. For inclined events the maximum is very distant from the impact point, effectively placing an upper limit on the zenith angle of $\sim$73$^{\circ}$ for both to be observable. For these events, the energy resolution is estimated as 12\%, at the highest energies, from a comparison of $N\mathrm{_{19}}$ with $E\mathrm{_{FD}}$ (Pierre Auger Collaboration, in preparation)\\

\noindent
For hybrid events, two estimates of the energy are available, namely that from the one, or more, fluorescence measurements, and that from the determination of \textit{S}(1000) and the use of the calibration data. For consistency, the latter value has been quoted in all cases as it is available for all events. Average uncertainties in energy of 8\% for vertical events and 12\% for inclined events are given. The systematic uncertainty in the energy estimates coming from those in \textit{S}(1000) depend on the distance spread of the signals in an event and on the presence, or otherwise, of saturated stations. The dominant systematic uncertainty in the energy estimates of 14\% comes from the FD measurements.

\clearpage
\section{The Events of the Catalog} \label{sec:catalog}
The catalog presented in this paper contains details of the 100 highest energy events recorded using the array of water-Cherenkov detectors of the Pierre Auger Observatory, together with similar data for a further nine events used in the energy-calibration procedures outlined in Section~\ref{sec:energyReco}. Full details of all 109 events are available at \url{https://opendata.auger.org/catalog/}. A list summarising the events is also included there. In this section, features of eight exemplary events are discussed in some detail to enable features in the full set of data to be appreciated. One of the two hybrid events discussed below has an energy lying just outside of the range of the top 100.\\

\noindent
The events are identified with a catalog number (\#$N$) that can be used to locate it in the depository, and by a name, PAOddmmyy, that indicates the day, month and year of detection.

\subsection{Description of Individual Events} \label{sec:events}

\subsubsection{Vertical Events}\label{sec:vertcalEvents}
\subsubsection*{}
\noindent

\begin{figure}[ht!]
 \epsscale{0.9}
\plotone{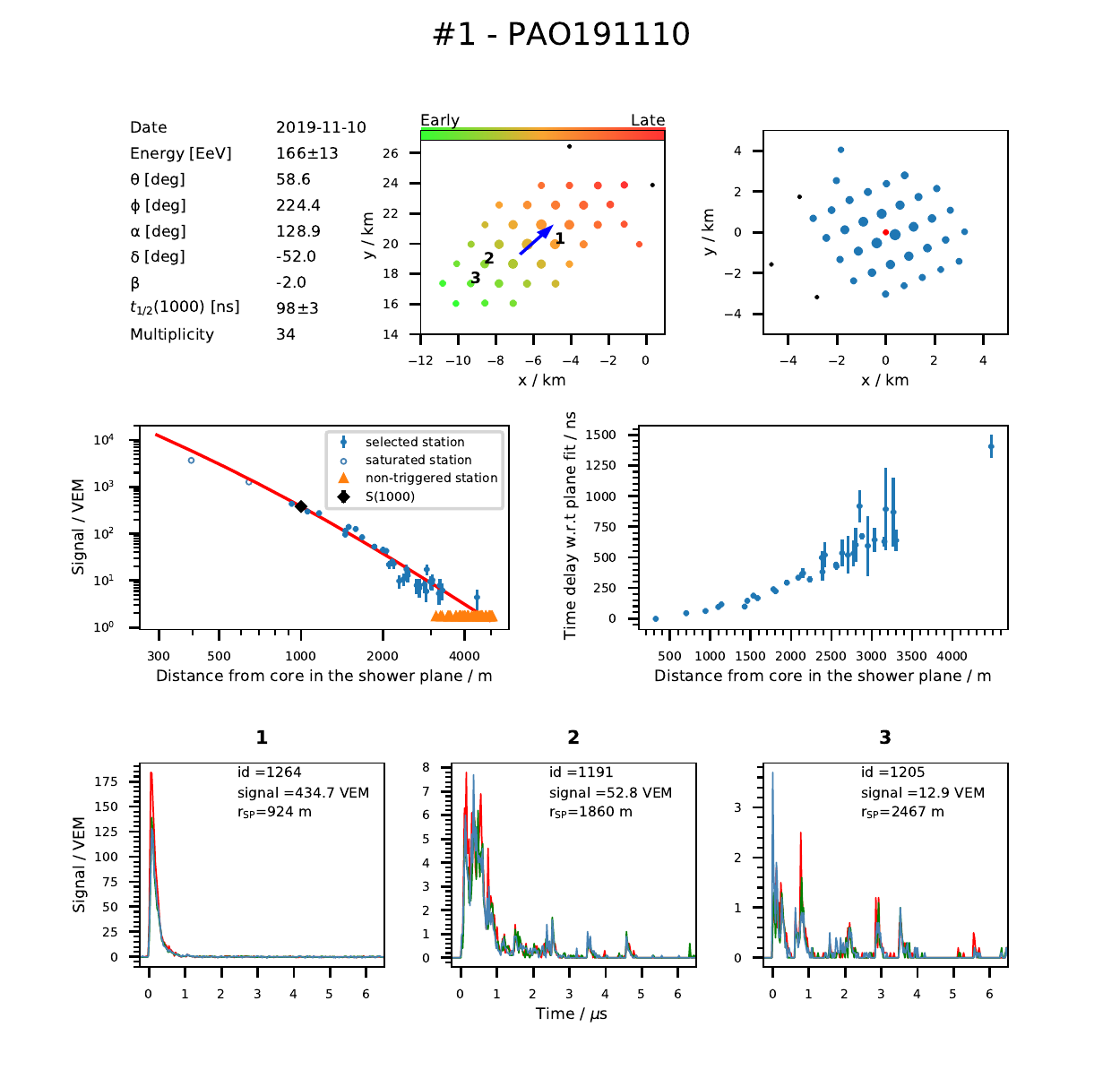}
\caption{Features of the most energetic event, (PAO191110, \#1) recorded with the Surface Detector of the Pierre Auger Observatory. See text for details.}
\label{fig:PAO191110}
\end{figure}

\textbf{PAO191110 (\#1):} Some properties of the most energetic air-shower registered with the Surface
Detector are shown in Figure~\ref{fig:PAO191110}. The primary energy is (166 $\pm$ 13)\,EeV with the shower impacting
the surface array at a zenith angle $\theta$ of 58.6$^{\circ}$. It has a right ascension $\alpha$ of 128.9$^{\circ}$ and a declination $\delta$ of $-52.0^{\circ}$. The top-middle panel shows the event footprint on the ground, which spans an area
of approximately (13 $\times$ 6)\,$\mathrm{km^2}$, with 34 water-Cherenkov detectors (WCDs) triggered. Black dots
correspond stations that triggered randomly. The detectors struck are shown in a plane perpendicular
to the direction of arrival in the top right-hand panel, where the red point corresponds to the
position of the shower core. The color coding and the blue arrow show the direction of
propagation of the air shower, evolving from green for detectors that trigger early through to red
for those that are triggered later. The radius of each circle is proportional to $\log S$, where $S$ is the signal size measured in VEM.\\

\noindent
In the left of the middle panel, the lateral distribution of the recorded signals, as a function of the
distance to the shower core, is shown. The triggered (blue circles) and non-triggered stations
(orange triangles) are indicated. The event has two saturated stations (blue open circles) close
to the shower core. Events with two saturated detectors are rare occurrences: only three events
in the full data sample have two detectors that are saturated simultaneously. The lateral spread
of the signals is described by the modified Nishimura-Kamata-Greisen (NKG) lateral distribution
function (LDF) discussed in Section~\ref{sec:recoIntro}. The value of the exponent $\beta$ in the LDF is given in the top-left panel. In the right of the middle panel, the time delays with respect to a fit that assumes a plane shower front is shown for the triggered stations. The delays are measured in ns.\\

\noindent
In the bottom three panels, the arrival time distributions of the signals recorded at three detectors (marked 1 to 3 on the signal map) are displayed. The different colors indicate the signals from the three photomultipliers in each detector. These traces exemplify how signal shapes vary with respect to the distance from the shower core. Here, and below, detectors have been selected that lie close to the distance (1000\,m) used to define the shower size (Section~\ref{sec:verticalReco}), and at other distances, selected according to the features being illustrated. It is known from direct measurements (\cite{Linsley1962}) that, except within a few meters of the shower axis, muons precede the electromagnetic component. The arrival times of the two components overlap to some extent, but the electromagnetic component lags the muon signals by an amount that increases with distance from the shower core. At 1000\,m, the risetime, ${t_{1/2}}(1000)$, in this event is close to 100\,ns (Section~\ref{sec:arrival}). 
The muons that are detected are typically minimum ionizing particles: as a result their signals show a fast risetime and a decay time that confines the signals over one to three 25\,ns time bins. As the distance to the shower core increases, there is more dispersion of the shower particles, with smaller signals that are spread out in time.\\

\clearpage
\subsubsection*{}

\begin{figure}[ht!]
 \epsscale{0.9}
\plotone{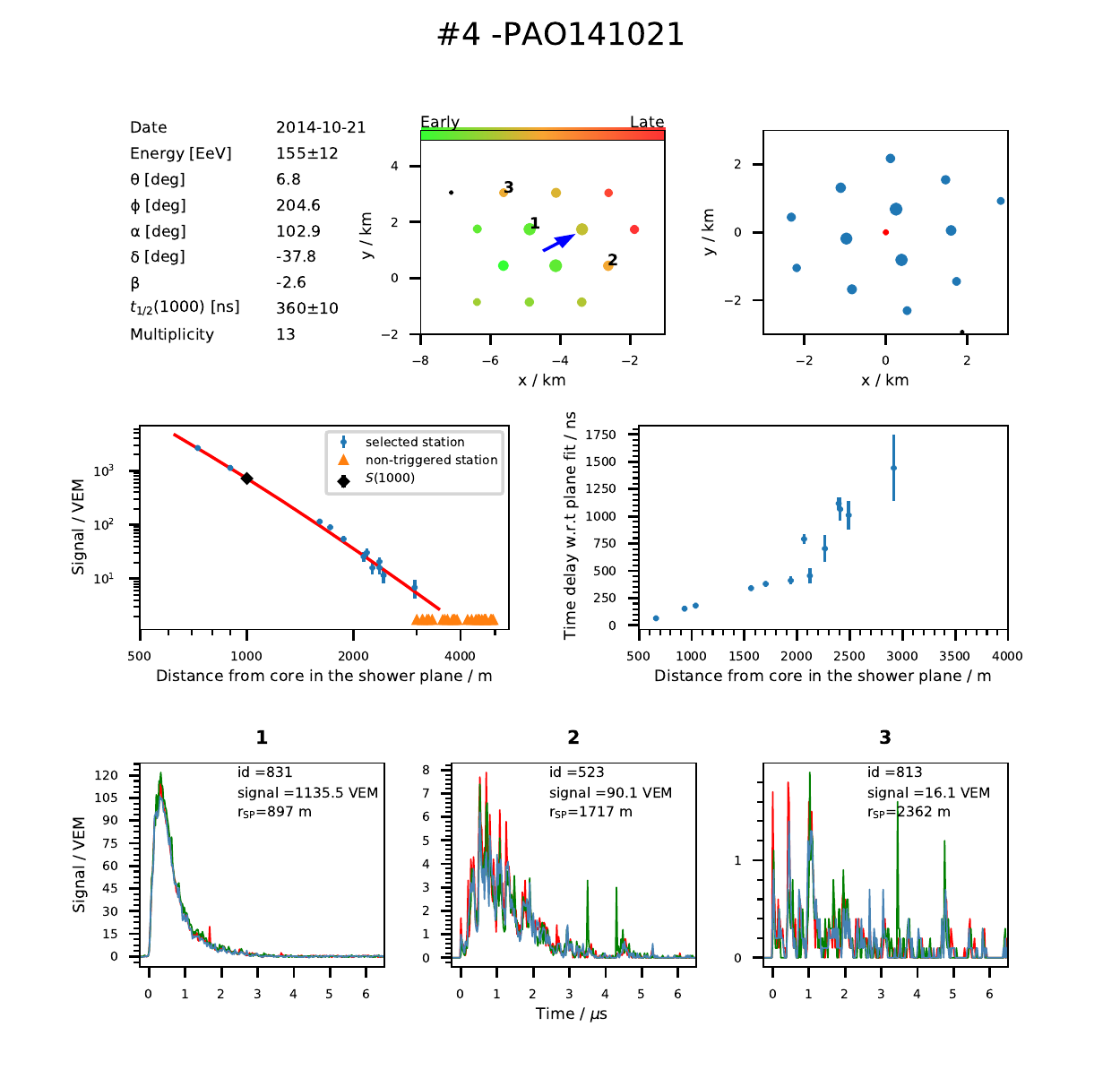}
\caption{Features of a vertical event with a reconstructed energy of 155\,EeV, (PAO141021, \#4). See text for details.}
\label{fig:PAO141021}
\end{figure}

\textbf{PAO141021 (\#4):} An event of primary energy (155 $\pm$ 12)\,EeV arriving at the ground-level at quasi-normal incidence (the measured zenith angle is 6.8$^{\circ}$) is shown in Figure~\ref{fig:PAO141021}. The footprint of the event is more compact and less elongated than that of PAO191110, (\#1). The top-middle panel shows the footprint on the ground, which spans an area of approximately (6 $\times$ 3)\,km$^2$: 13 WCDs are triggered. The middle panels show the lateral distribution of the recorded signals as a function of the distance to the shower core on the left, and on the right, the time delays with respect to a plane shower front, perpendicular to the incoming direction of the shower.

The signals and arrival times (lower panels) of the particles recorded at the three selected detectors are markedly different from those selected for event PAO191110, (\#1). The station with the largest signal (at $897\,$m from the core) is above 1000\,VEM, a factor of 2.6 greater than the signal in event PAO191110 recorded at a similar distance, 924\,m, from the core. As the distance travelled through the atmosphere is substantially shorter for this near-vertical event, the particles suffer less attenuation, resulting in a larger contribution to the signal from the electromagnetic component. This is reflected in the slower risetime: $t_{1/2}(1000) = (360~\pm~10)\,\text{ns}$. 
 Likewise, a $\beta$ value of $-2.6$ indicates that the LDF of this event is steeper than that of event PAO191110 for which $\beta$ is $-2.0$.

\clearpage
\subsubsection*{}
\noindent

\begin{figure}[ht!]
 \epsscale{0.9}
\plotone{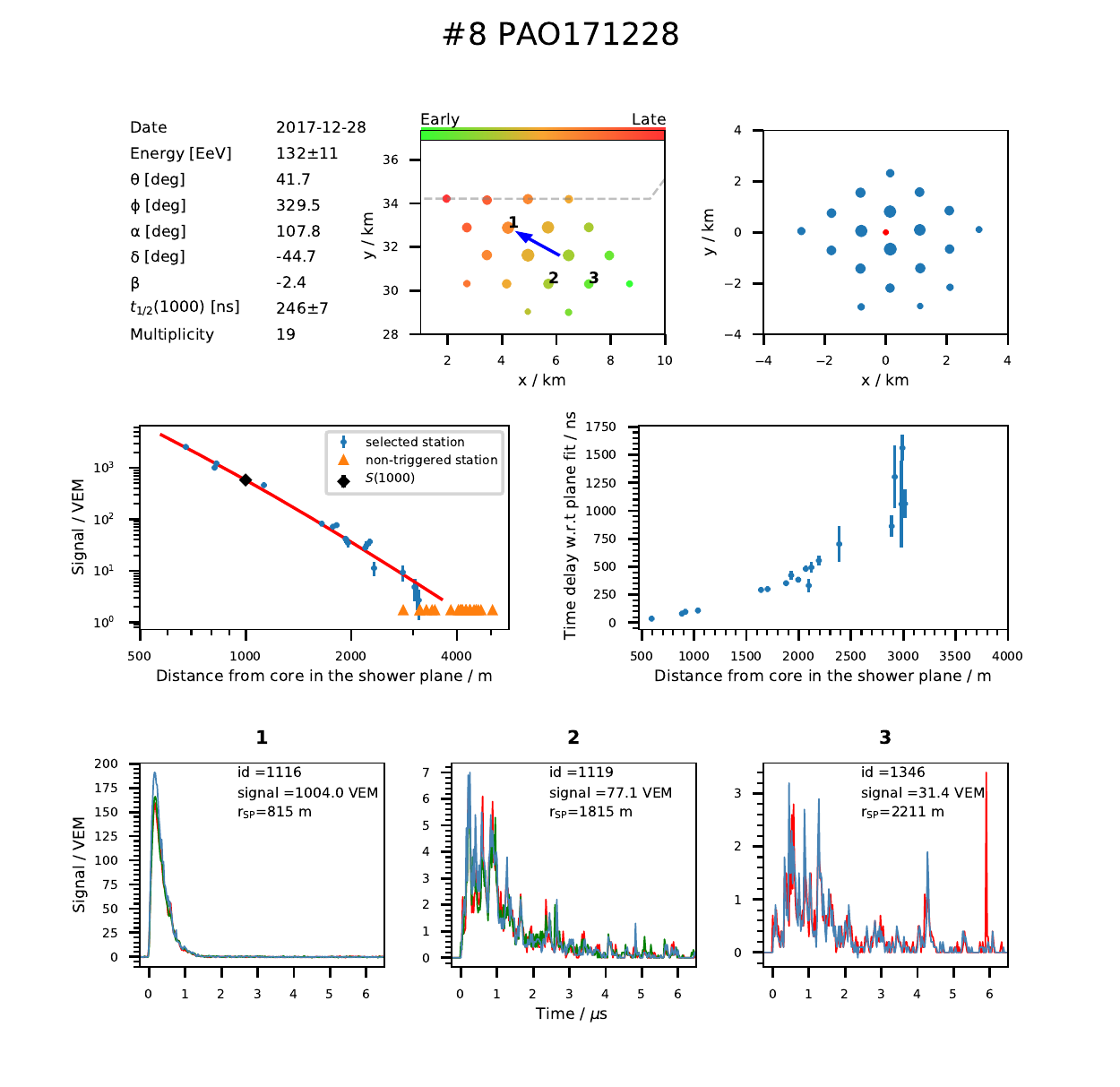}
\caption{Features of an event, (PAO171228, \#8) recorded with WCDs located at one of the boundaries of the Surface Detector. See text for details. Note that in detector \#1346 only two of the three photomultipliers were operational. Reality dictates that it is impossible to keep all three photomultipliers operational 100\% of the time. Failures of two, or even all three, photomultipliers inevitably occur. Typically 98\% of all stations are active at any time, sending triggers at 20\,Hz to the central station (Section~\ref{sec:recording}). }
\label{fig:PAO171228}
\end{figure}

\textbf{PAO171228 (\#8):} An event with primary energy (132 $\pm$ 11)\,EeV arriving with zenith angle $\theta = 41.7^{\circ}$ shown in Figure~\ref{fig:PAO171228}. As can be seen in the top-middle panel, only 19 WCDs have been triggered
because the footprint of this event extends beyond the limits of the array (the dashed grey line
marks the perimeter). Although the event is not fully contained, the reconstruction of the main
observables used in the various physics analyses (Section~\ref{sec:arrival}) is of high-quality.\\

\noindent
In the bottom-right panel (station id \#1346) there is a signal of over 3\,VEM at about $6\,\upmu$s. Such
signals are due to a contribution from direct light reaching one photomultiplier and are likely
caused by the passage of a particle close to the location of the photomultiplier, perhaps moving
in an upward direction, or possibly due to light from an electron produced in a muon decay where
the decay electron has been emitted towards the photomultiplier. Under these conditions, the
Cherenkov photons are detected directly, and a sharp, distinctive signal is recorded by a single
photomultiplier, rather than the broader signals produced when the light is scattered on the
inner reflective walls of the WCDs. The increase in signal size caused by the direct light varies with distance and is typically about 1\% at 1000\,m for events arriving close to the vertical.

\clearpage
\subsubsection*{}
\begin{figure}[ht!]
 \epsscale{0.9}
\plotone{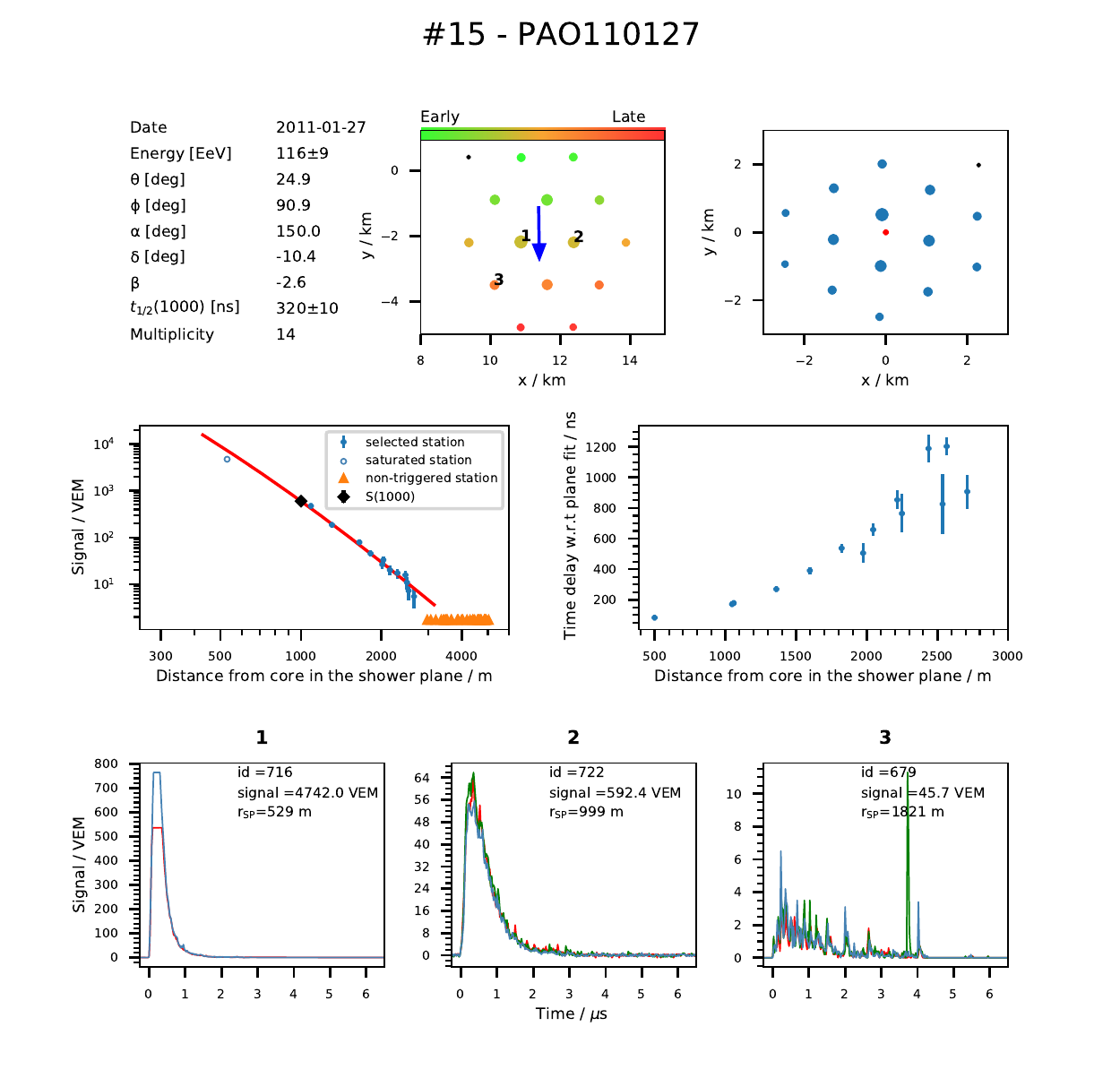}
\caption{Features of an event, (PAO110127, \#15) in which one of the WCDs is saturated. See text for details.}
\label{fig:PAO110127}
\end{figure}

\
\textbf{PAO110127 (\#15)}: This event (Figure~\ref{fig:PAO110127}) has been selected to show some singular signals that are
relatively rare. In this event 14 water-Cherenkov detectors were triggered and used to measure
the energy, (116~$\pm$~9)\,EeV, zenith angle $\theta = 24.9^{\circ}$, and risetime at 1000\,m being (320~${\pm}$~10)\,ns. 
However, the detector closest to the core (located at just over 500\,m) shows a saturated signal (see the bottom-left panel in the figure). 
In this case, the saturation is due to the overflow of the finite dynamic
range of the read-out electronics. The procedure used to recover the majority of such signals is
discussed in Section~\ref{sec:verticalReco} above.\\

\noindent
The bottom-right panel (station id \#1346) again exemplifies, as in Figure~\ref{fig:PAO171228}, a signal of over 10\,VEM at about 3.8\,$\upmu$s that contains a contribution from direct light reaching one photomultiplier.

\newpage
\subsubsection{Inclined Events}\label{sec:inclinedEvents}

\subsubsection*{}
\noindent

\begin{figure}[ht!]
 \epsscale{0.9}
\plotone{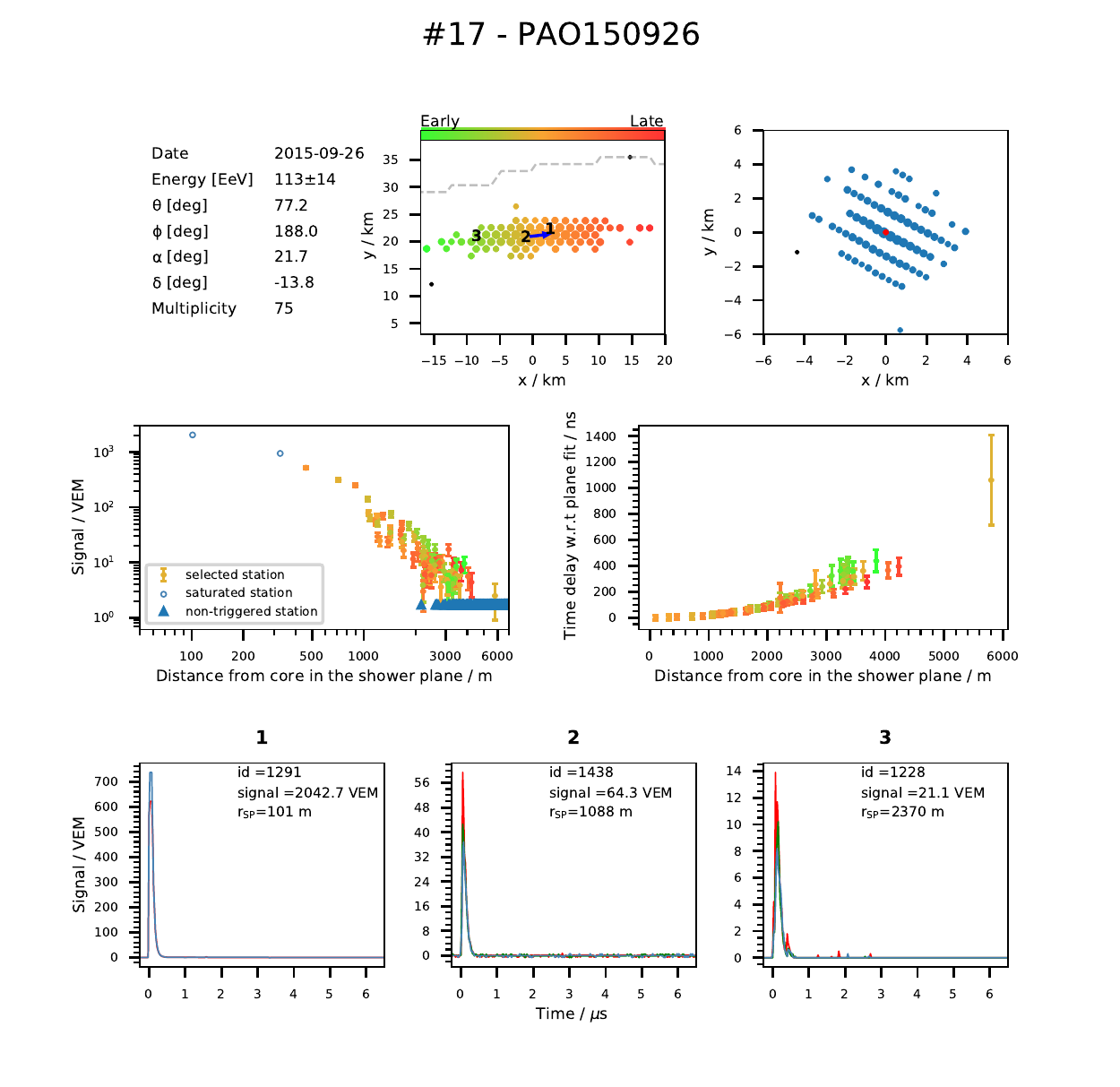}
\caption{Features of the most energetic event, (PAO150926, \#17) belonging to the set of inclined showers ($\theta >~60^{\circ}$).
See text for details.}
\label{fig:PAO150926}
\end{figure}

\textbf{PAO150926 (\#17):} The inclined event, zenith angle $\theta = 77.2^{\circ}$, with the highest energy, (113~$\pm$ 14)\,EeV, is shown in Figure~\ref{fig:PAO150926}. The shower triggered 75 WCDs in an elongated pattern on the ground, over an area close to (35 $\times$ 6)\,km$\mathrm{^2}$. The shower particles must traverse long distances to reach the
ground at such inclinations. Thus, electromagnetic particles are mostly absorbed in the
atmosphere and the signals at the ground are produced almost entirely by muons. In contrast to
events with lower inclinations, most of the signal arrives within a very short time of around 200\,ns, independent of the location within the shower footprint (see bottom row in Figure~\ref{fig:PAO150926}).
Likewise, the distribution of the integrated signal on the ground loses the near-rotational
symmetry of more vertical events (Section~\ref{sec:recoIntro}). Hence, the distribution of the recorded signals as a function of the distance to the shower core shown in the left middle panel cannot be
described by a single rotationally-symmetric function. In the middle-right panel, the delay of the start of the signal in
each triggered WCD with respect to a plane shower front is presented. The shower is very
asymmetric and cannot be well described by, for example, a concentrically-inflated spherical model.\\

\clearpage

\begin{figure}[ht!]
 \epsscale{0.5}
\plotone{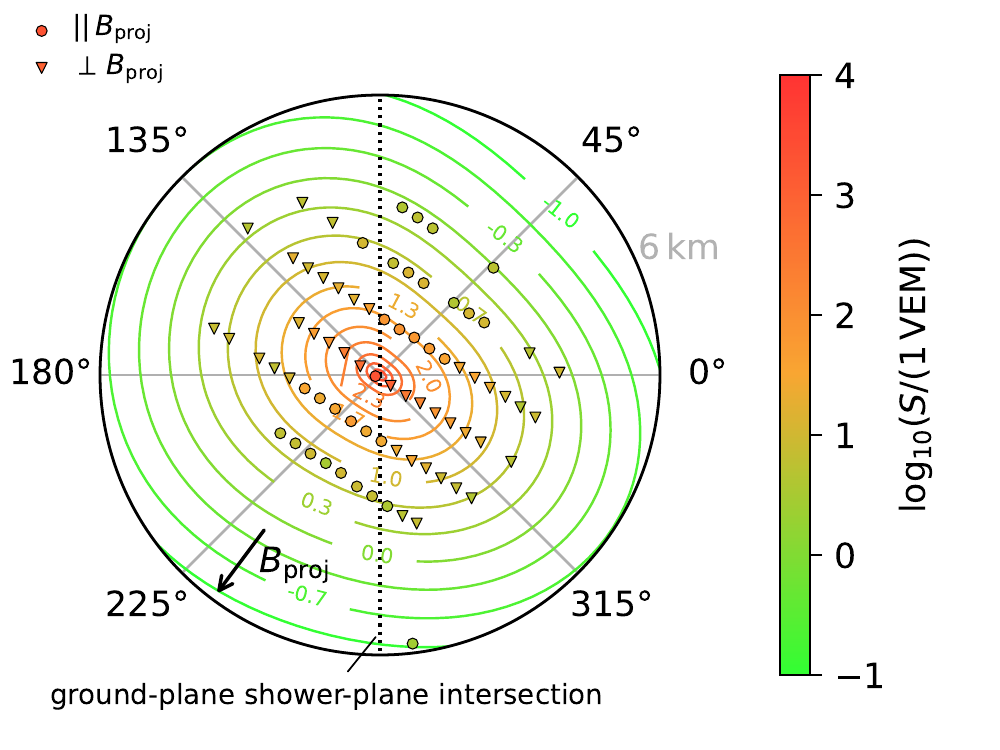}%
 \quad
\epsscale{0.6}
\plotone{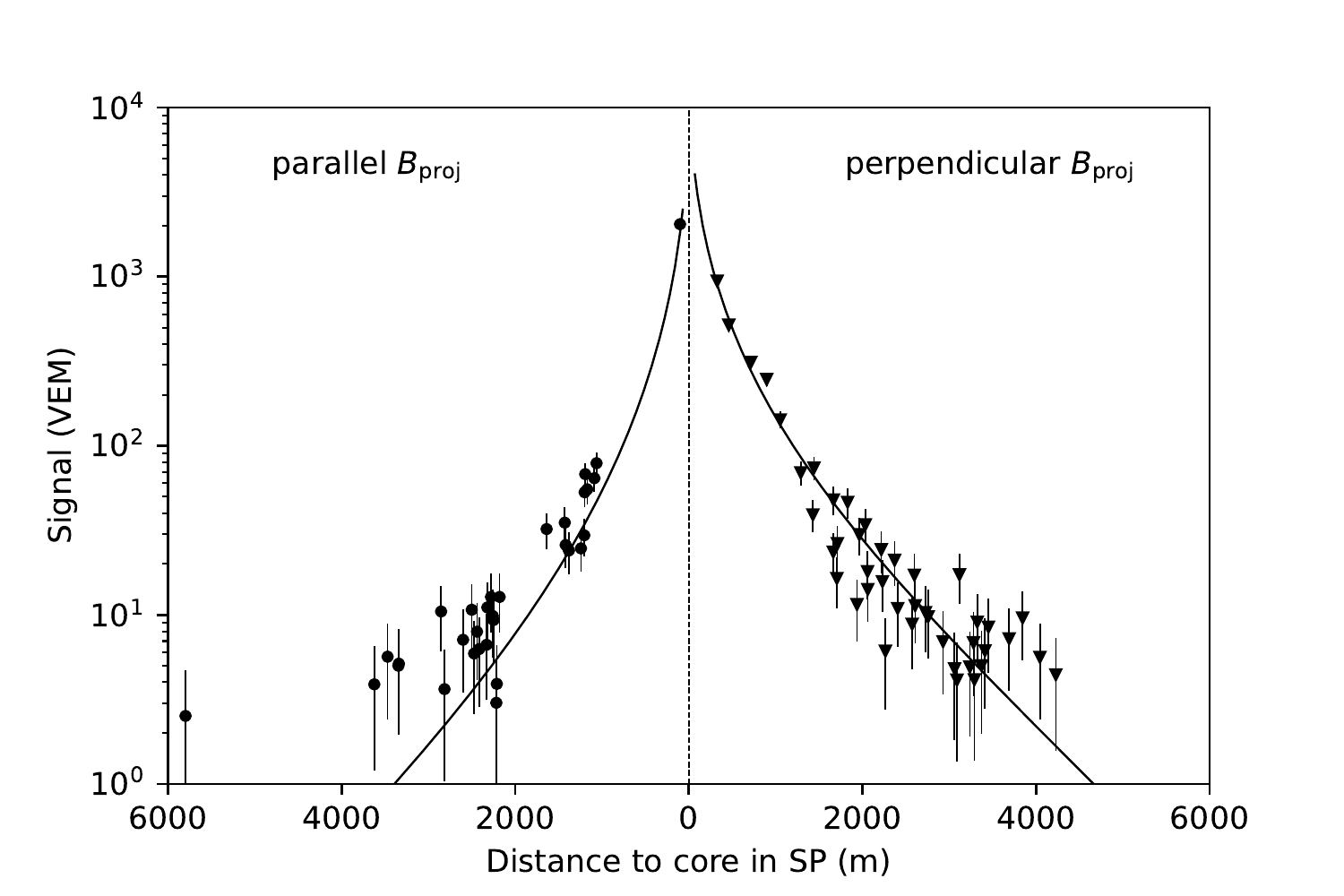}
\caption{
\emph{Left:} 2D distribution of measured and expected signals in the shower plane. Station markers are colored according to the signal. The direction of the magnetic field in the shower plane ($B\mathrm{_{proj}}$) is indicated by the black arrow. Triangular markers are stations which, seen from the position of the core, lie within $\pm$ 45$^{\circ}$ of a direction perpendicular to $B\mathrm{_{proj}}$. That is, these stations are in the direction of the deflection that charged particles experience in the magnetic field. More particles therefore reach those stations (enhancing the signal) compared to stations that are at the same distance to the core but that lie along the direction of the magnetic field (circular markers). The intersection of the shower plane with the ground plane is shown by the dashed line. 
\emph{Right:} projection of the signal distributions as a function of the distance from the shower core. The markers show the signal measured at the stations, while the curves show the expected signal. Stations in the direction parallel to the magnetic field are shown on the left, with stations in the direction perpendicular to the magnetic field on the right. 
}
\label{fig:inclined1}
\end{figure}

The reconstruction, using a 2-dimensional pattern of muon densities at the ground (Section~\ref{sec:inclinedReco}) for this event, is presented in Figure~\ref{fig:inclined1}. 
In the left panel, the distribution of the triggered stations around the shower core in the plane perpendicular to the shower direction (the shower plane) is shown in polar coordinates. The coordinate system is such that the y-axis coincides with the intersection of the ground plane with the shower plane (dashed line). Polar angles close to zero (along the positive x-axis) correspond to stations triggering before the shower core arrives at the ground (so-called ‘early stations’), while angles towards 180$^{\circ}$ correspond to ‘late stations’. The colored contour lines indicate the expected signal for the distribution of muon densities that best fits the observed signals. The direction of the component of Earth’s magnetic field in the shower plane is indicated by the black arrow. Note how the signal pattern is distorted in the direction perpendicular to the magnetic field. In addition to the distortion induced by the geomagnetic field, there is a small difference between the signals of early (right of dashed line) and late stations (left of dashed line). This difference arises from the attenuation of muons, and also from the different angles of incidence of muons on the detectors. In the right-hand panel slices of the LDF parallel and perpendicular to the projected magnetic field are shown.

\clearpage
\subsubsection*{}

\begin{figure}[ht!]
 \epsscale{0.9}
\plotone{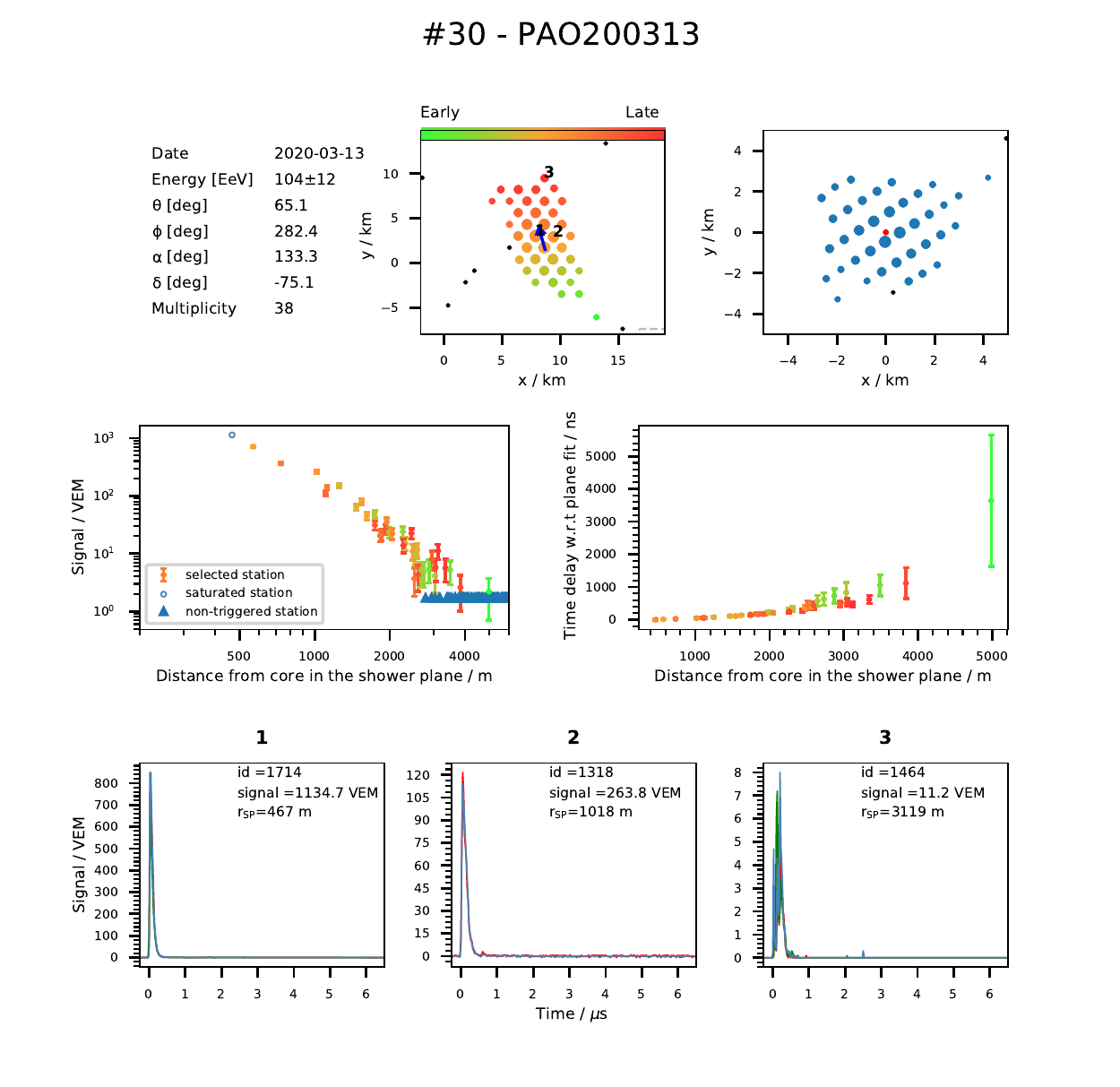}
\caption{Features of the second most energetic event, (PAO200330, \#30) belonging to the set of inclined showers ($\theta > 60^{\circ}$). See text for details.}
\label{fig:PAO200313}
\end{figure}

\begin{figure}[ht!]
 \epsscale{0.5}
\plotone{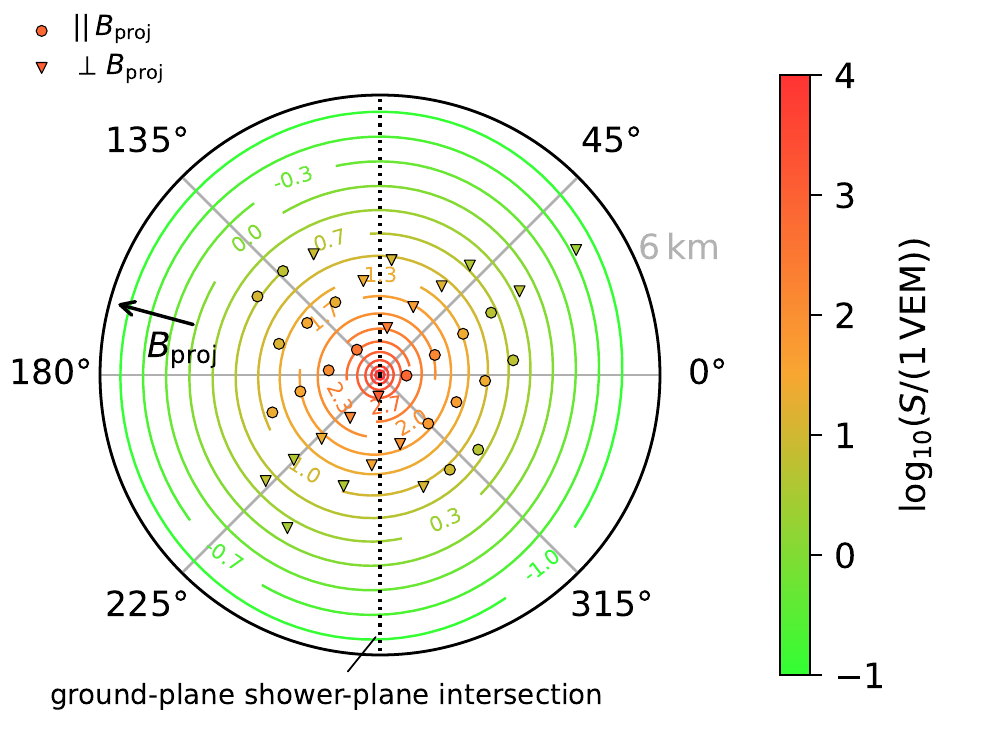}%
 \quad
\epsscale{0.6}
\plotone{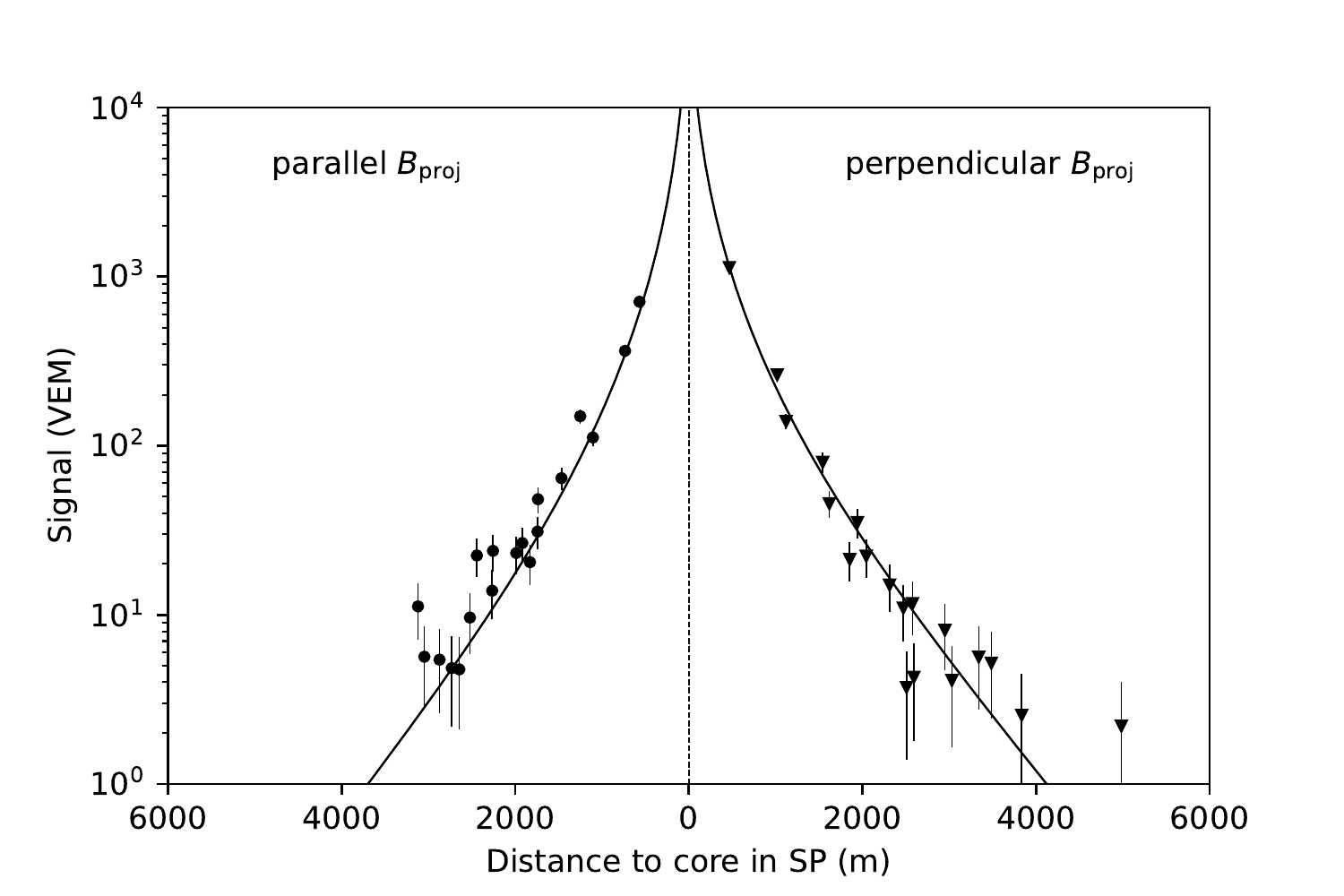}
\caption{
 \emph{Left:} 2D distribution of measured and expected signals in the shower plane. 
\emph{Right:} the projection of the signal distribution onto the distance from the shower core. Note the slight asymmetry between the left (late stations) and right half (early stations) of the signal in the shower plane (left panel) and the lack of asymmetry between stations parallel and perpendicular to the magnetic field (right panel).}
\label{fig:inclined2}
\end{figure}

\textbf{PAO200313 (\#30):} 
This event (Figure~\ref{fig:PAO200313}) is the second highest-energy inclined event with an energy of (104~$\pm$~12)\,EeV. At a zenith angle of $\theta = 65.1^{\circ}$, this shower triggered 38 detector stations in an elongated pattern on the ground (19~$\times$~6)\,km$\mathrm{^2}$. As in the previous case, the shower pattern at the ground shows some asymmetry. Even at this inclination, there is a substantial electromagnetic component present and an additional 3\,km of atmosphere (the early-late effect) corresponds to more than five radiation lengths. Thus, the asymmetry arises dominantly from the difference in the attenuation of the electromagnetic component rather than from deflections of the muons in the geomagnetic field. The effect is illustrated in Figure~\ref{fig:inclined2}.

\clearpage

\subsubsection{Hybrid events}\label{sec:hybridEvents}
The first of the two events discussed here passes the high-quality criteria applied to select the sub-sample of hybrid events used for energy calibration (Section~\ref{sec:energyReco}) of vertical events. The second event represents the most energetic shower used in the calibration of inclined events. The details of the ten most energetic hybrid events used for calibration, including those described below, can be found at \url{https://opendata.auger.org/catalog/}.

\subsubsection*{}

\begin{figure}[ht!]
 \epsscale{0.9}
\plotone{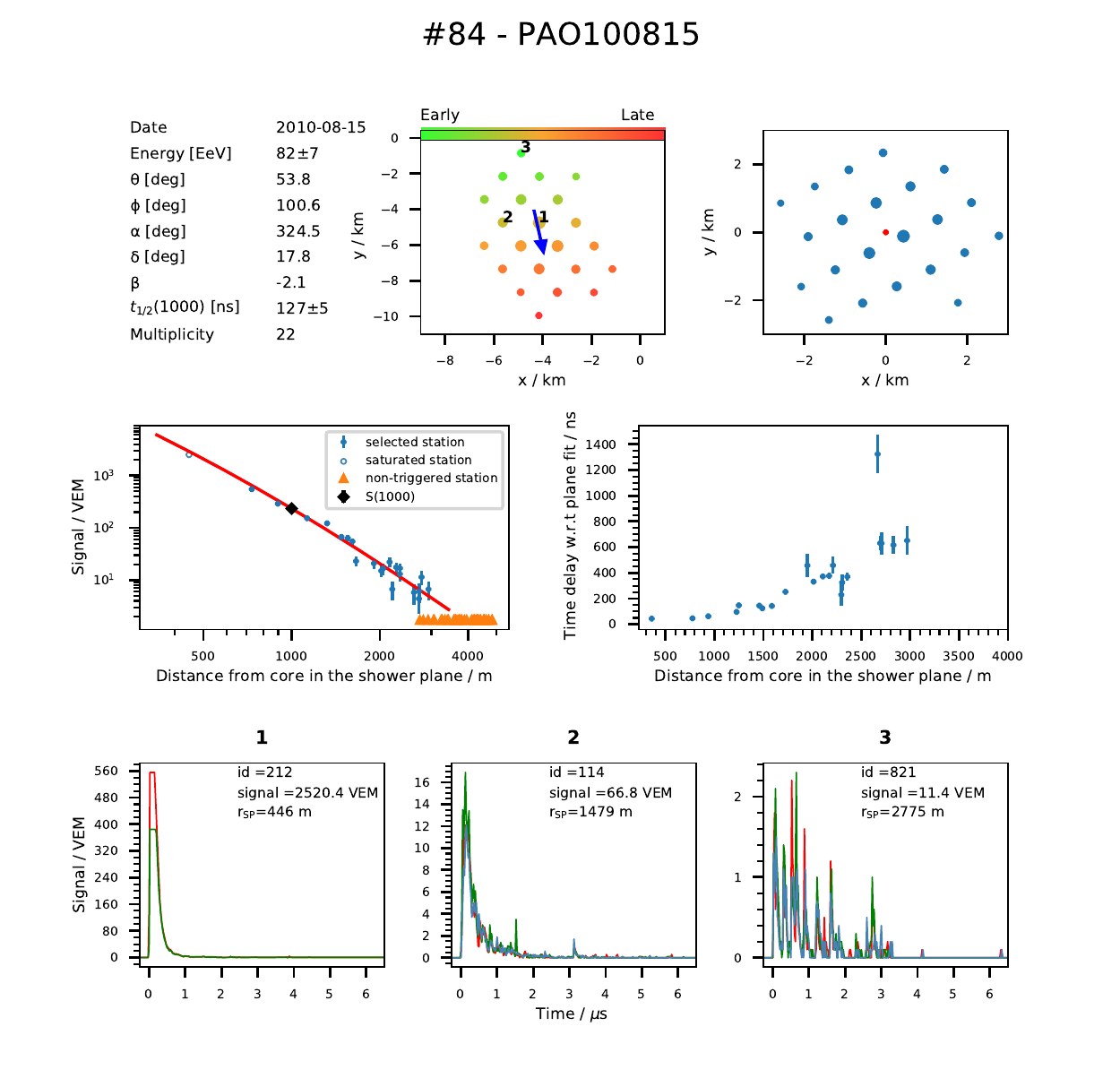}
\caption{Features of the most energetic hybrid event, PAO100815, \#84. See text for details.}
\label{fig:PAO100815_1}
\end{figure}

\begin{figure}[ht!]
 \epsscale{1}
\plotone{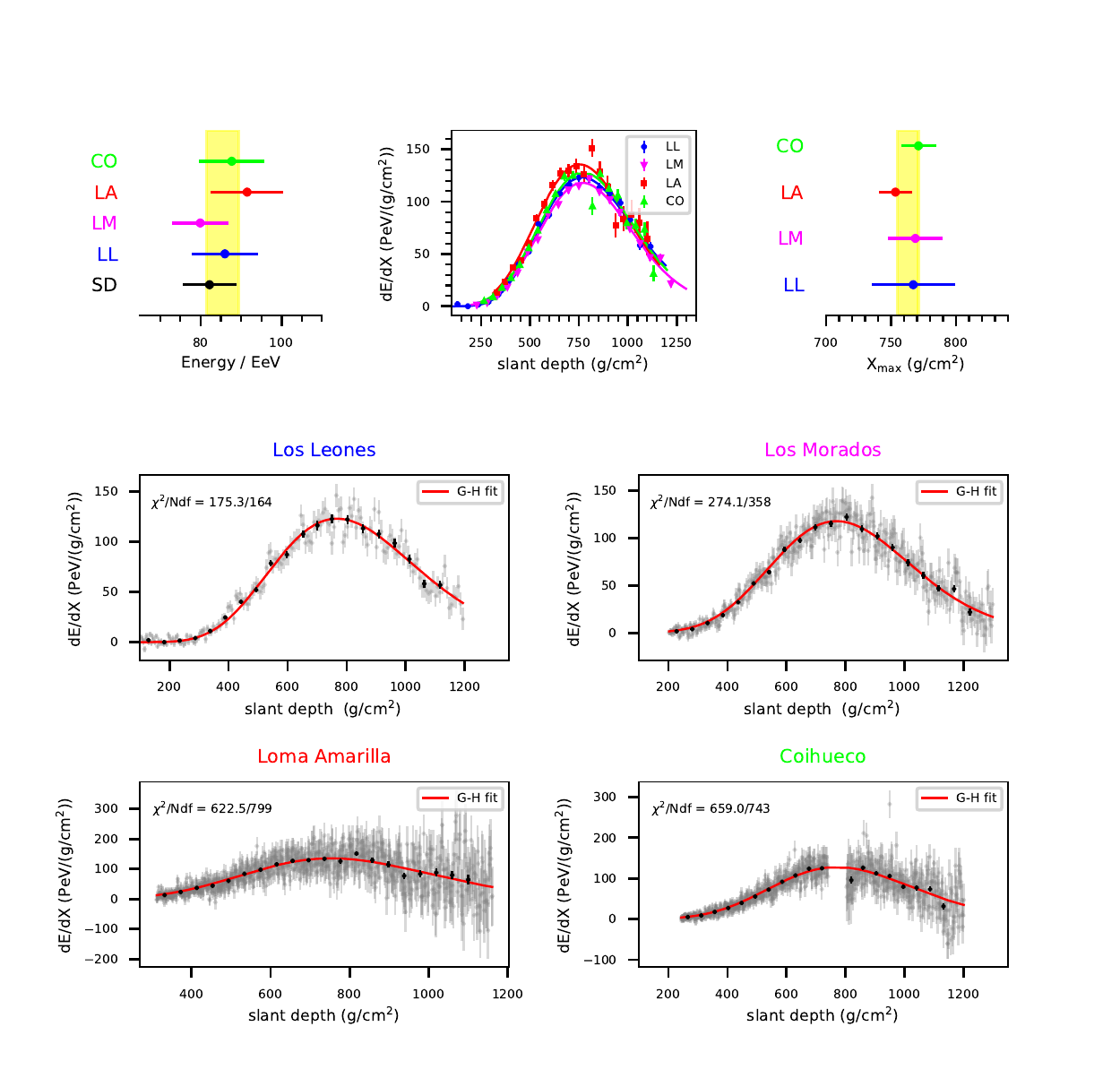}
\caption{
Reconstructed parameters of PAO100815, \#84. FD stations used in the reconstruction are distinguished by different colors. The red lines correspond to fits to the profiles of the energy deposition using the universal shower profile function (Section~\ref{sec:FDreco}). The yellow bands are centered on the combined weighted average of the measurements of $X\mathrm{_{max}}$ and the energy at the FD sites. The widths of the bands correspond to the statistical uncertainties of combinations. The uncertainty in the SD energy is 8\% (Section~\ref{sec:energyReco}).}

\label{fig:PAO100815_2}
\end{figure}

\begin{figure}[ht!]
% \epsscale{0.9}
 \centering
\includegraphics[trim={0 2.5cm 0 0}]{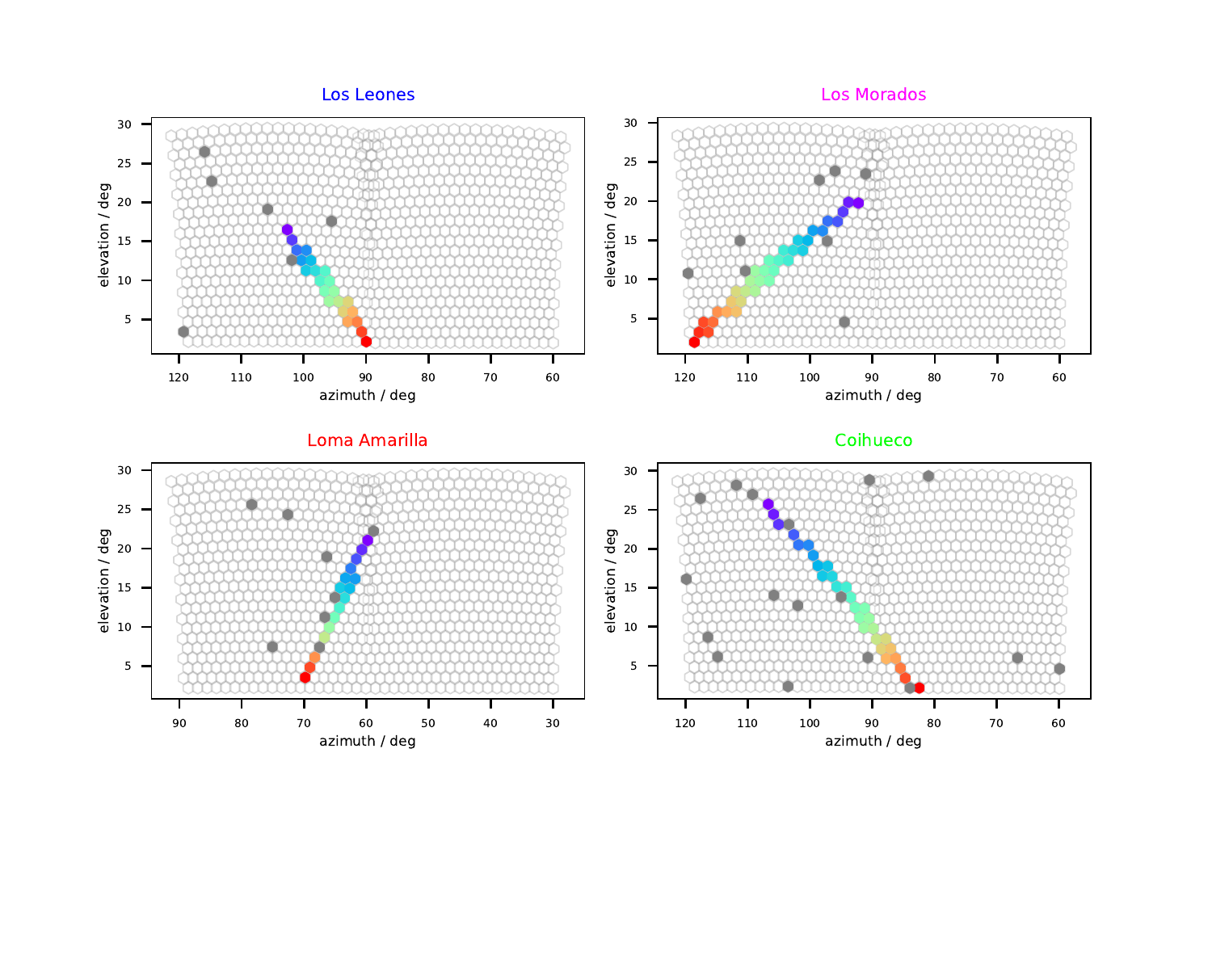}
\caption{The camera views in all four telescopes for event PAO100815, \#84. The colors (violet to red) indicate the times (early to late) at which the light reaches each pixel. Dark pixels are random coincidences and not used in the reconstruction.}
\label{fig:PAO100815_3}
\end{figure}

\begin{figure}[h]
 \epsscale{0.7}
\plotone{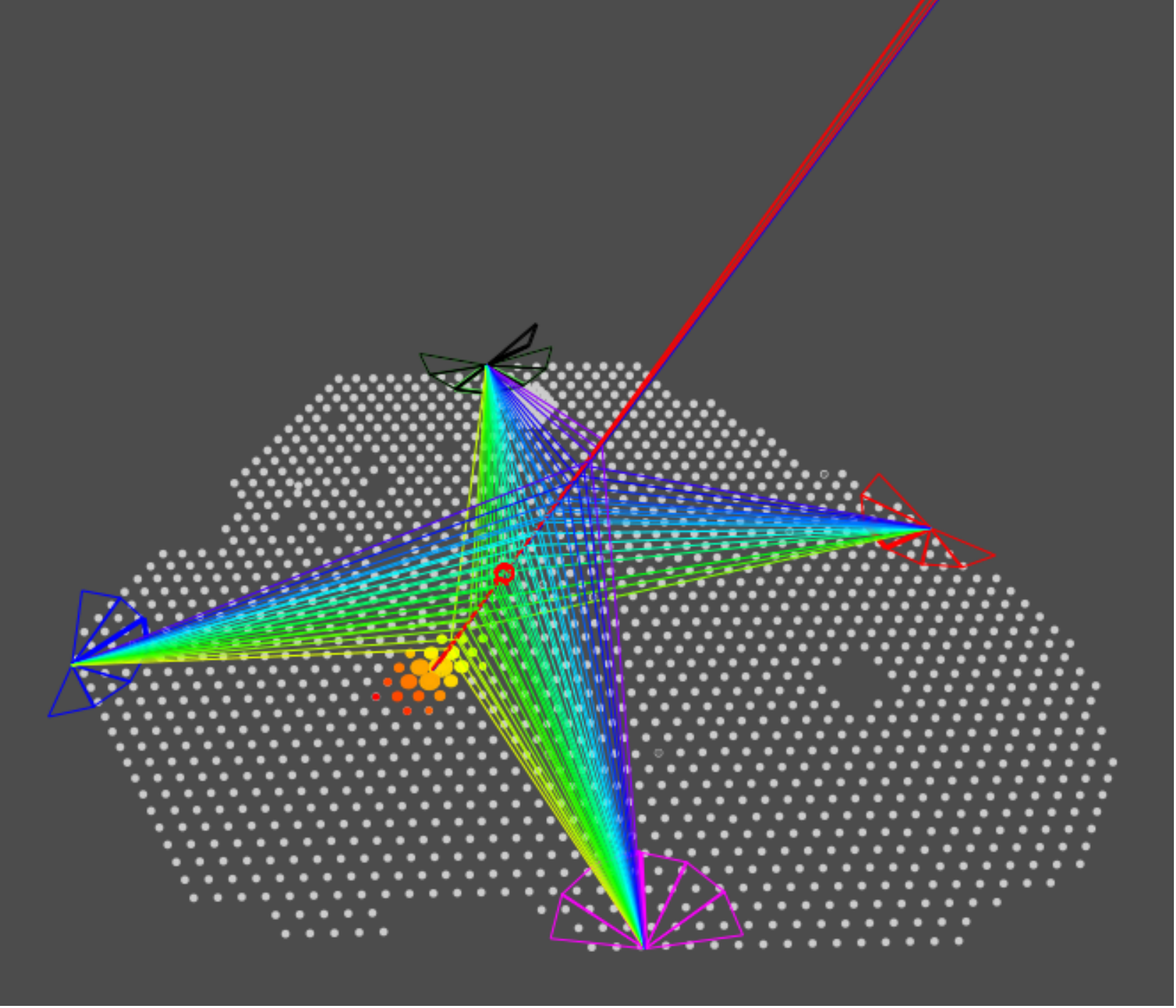}
%\centering
%\includegraphics[trim={2cm 2.5cm 0 0},width=0.6\textwidth]{3Devent_1.pdf}
\caption{Three-dimensional visualization of PAO100815, \#84. The lines correspond to the light rays and point to the
telescopes of the fluorescence detectors. The colors of the light rays and of the SD stations represent trigger times
of FD and SD PMTs, respectively.}
\label{fig:3Devent}
\end{figure}

\textbf{PAO100815 (\#84):} 
This is the most energetic hybrid event, arriving at a zenith angle $\theta = 53.8^{\circ}$. Details of the event are shown in Figures~\ref{fig:PAO100815_1}~to~\ref{fig:3Devent}. The energy estimate from the determination of \textit{S}(1000) is (82~$\pm$~7)\,EeV, consistent with that from the fluorescence measurements of (85~$\pm$~4)\,EeV. There are 22 triggered stations with a footprint of about (7.5 $\times$ 6)\,km${^2}$. The lateral distribution of signals is described by the modified NKG function. The signals registered by the WCDs are shown in the bottom panels of Figure~\ref{fig:PAO100815_1}. The light received at the station about 450\,m from the shower core (left panel) has saturated the dynamic range of the two photomultipliers (see Section~\ref{sec:energyReco} and event PAO110127, \#15 above) that were operational. The amplitude difference indicates the complexity of the saturation process. For the two detectors with distances to the core larger than 1000\,m, the FADC show the typical structure of shower signals, where the early parts of the FADC traces are dominated by muons and the tails are populated with broader signals due to photons, electrons and positrons. The risetime at 1000\,m is (127~$\pm$~5)\,ns.\\

\noindent
Fluorescence light was detected at all four FD stations. Each individual hybrid-reconstruction passed the selection criteria. The reconstructed profiles of the energy deposition in the atmosphere are shown in the lower part of Figure~\ref{fig:PAO100815_2}, while the reconstructed energies (Section~\ref{sec:energyReco}) and depths of shower maximum ($X\mathrm{_{max}}$) are displayed in the upper section of the figure. \\

\noindent
Shower events crossing the field of view of a telescope at larger distances have lower angular velocities than those that pass close to the telescope. Additionally, when a shower is observed approaching the telescope, the signals are registered more rapidly across the camera than for
those from showers moving away from it. These effects result in different angular velocities of the shower images on the telescope cameras. Accordingly, the number of points is different in the profiles of the energy deposit recorded at the individual stations. The discrete binning of the energy deposits is a consequence of the 100\,ns readout of the photomultipliers of the fluorescence telescopes.\\

\noindent
The uncertainties in the energy and $X\mathrm{_{max}}$ estimates from individual stations of the Fluorescence Detector differ mainly because different amounts of Cherenkov light are detected at them. The relatively larger fraction of Cherenkov light (12\%) at the Los Leones station, results in a larger uncertainty in the longitudinal profile because Cherenkov emission is strongly beamed around the shower axis. Thus, a small uncertainty in the shower geometry translates into a larger uncertainty in this profile when compared with the estimate from Coihueco, where the Cherenkov light is only 5\% of the integral of the light flux. The uncertainty is also affected by other effects, such as the distance of the shower to the FD sites, that result in different numbers of photons being detected. At the Coihueco site, the shower image is detected at two telescopes, giving rise to a gap in the reconstruction of the profile of deposited energy. This occurs because the times for which the 
shower image is close to the border of the field of view of a telescope are rejected as it is not 
possible to make an accurate estimate of the light flux. Overall, the $X\mathrm{_{max}}$ and energy estimates from individual FD stations agree within quoted statistical uncertainties.\\

\noindent
In Figure~\ref{fig:PAO100815_3}, the camera views are shown for all eight telescopes at the four sites where the event was detected. The colors assigned to individual pixels represent centroids of pulses in the photomultipliers, thus marking the arrival time of fluorescence and Cherenkov light at the telescopes. Dark grey pixels indicate pixels that triggered randomly that do not match the time fit used to determine shower geometry (Section~\ref{sec:FDreco}). These random triggers arise from the night-sky background that varies for each detected shower and with the direction in which a telescope is pointing. There are no such pixels in the telescopes shown in event PAO140131, \#101 (Figure~\ref{fig:PAO140131_2}). The horizontal axes in the camera views correspond to local azimuth angles, defined counter-clockwise from the back-wall of the FD station. The origin points to the right, looking on to the shower from the position of the station. 
The vertical axis is an angular elevation of the viewing direction of the FD pixels. 

\clearpage
\clearpage

In Figure~\ref{fig:3Devent} a three-dimensional view of the event is exhibited.

\newpage
\subsubsection*{}
\begin{figure}[ht!]
%\epsscale{0.6}
\plotone{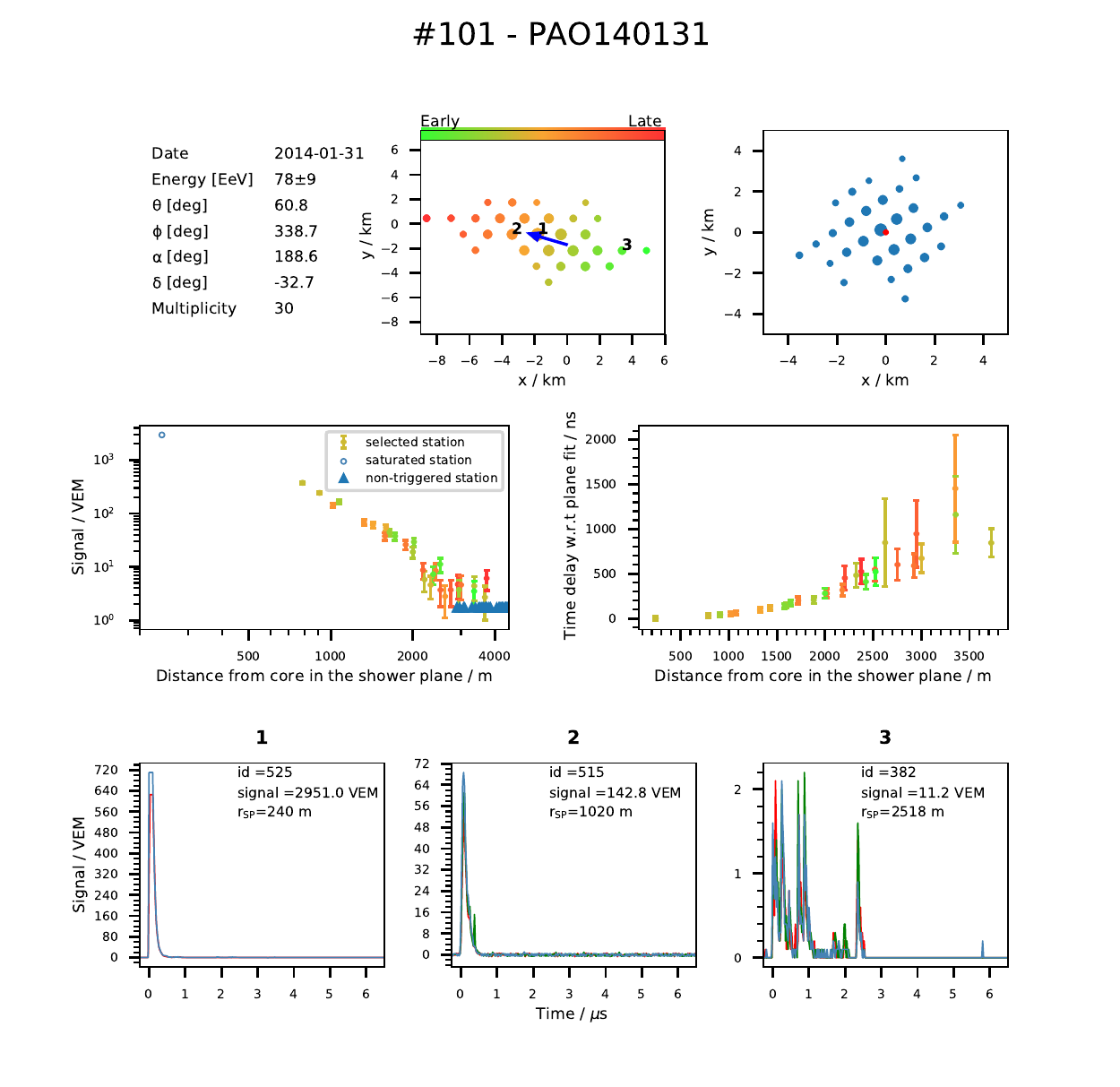}
\caption{The parameters reconstructed using the data from the WCDs for event, PAO140131, \#101.}
\label{fig:PAO140131_1}
\end{figure}

\
\textbf{PAO140131 (\#101):} 
This is the second most energetic hybrid event and belongs to the dataset used to calibrate events with zenith angle above 60$^{\circ}$. The zenith angle $\theta = 60.8^{\circ}$. The energy reconstructed from the SD signals is (78~$\pm$~9)\,EeV, consistent with that from the fluorescence measurement of (73~$\pm$~8)\,EeV. With 30 triggered stations, the footprint is elongated and covers an area of ($14 \times 6$)\,km$^2$. At 60$^{\circ}$, the depth of the atmosphere is twice the atmospheric vertical depth. Thus the electromagnetic component of the shower is partially quenched (see Section~\ref{sec:verticalReco}). The lateral distribution function and the time delay of the start time signals are barely asymmetric (see Figure~\ref{fig:PAO140131_1}) and can thus be described by the modified NKG function used for the vertical reconstruction.\\

\begin{figure}[ht!]
%\epsscale{0.6}
\centering
\includegraphics{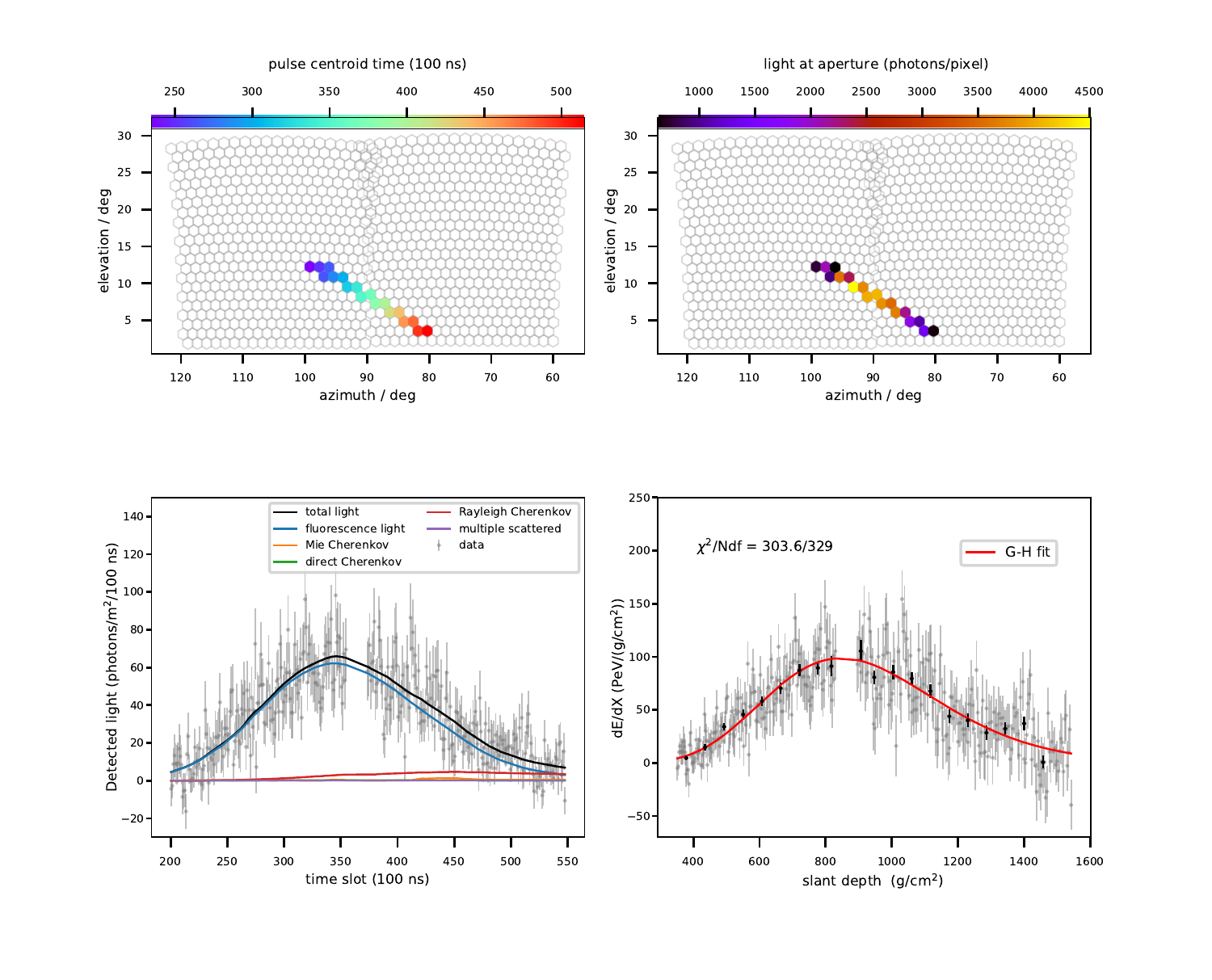}
\caption{Data from the fluorescence telescopes in event PAO140131, (\#101). The profile of the energy deposits
(bottom right) is accompanied by the light flux profile (bottom left) and camera views from two telescopes at Loma
Amarilla (top). The shower fell far from the telescopes with the closest point to the shower axis being about 35\,km.}
\label{fig:PAO140131_2}
\end{figure}

\noindent
Fluorescence light was detected at three FD stations (Los Morados, Coihueco and Loma Amarilla), but only the reconstruction for Loma Amarilla passed the selection criteria. The profile of energy deposition (Figure~\ref{fig:PAO140131_2}, bottom-right) is obtained from the profile of the detected light (Figure~\ref{fig:PAO140131_2}, bottom-left). The color bands in the figure of the light flux profile show the contributions from different light sources. Fluorescence light dominates, while Cherenkov light scattered into the telescope makes up 10\% of the integrated signal. \\

\clearpage

The top panels of Figure~\ref{fig:PAO140131_2} show the camera views of the shower crossing two adjacent telescopes at the Loma Amarilla site. The photomultipliers are sequentially triggered (top-left panel with colors coding the trigger time). The charges at each photomultiplier are proportional to the light flux received at the entrance window of each telescope. The shower image is detected in two telescopes giving rise to a gap in the reconstruction of the profile.

\clearpage
\section{A Sky Map of the 100 highest-energy events}\label{sec:skyMap}
\begin{figure}[ht!]
%\epsscale{0.6}
\plotone{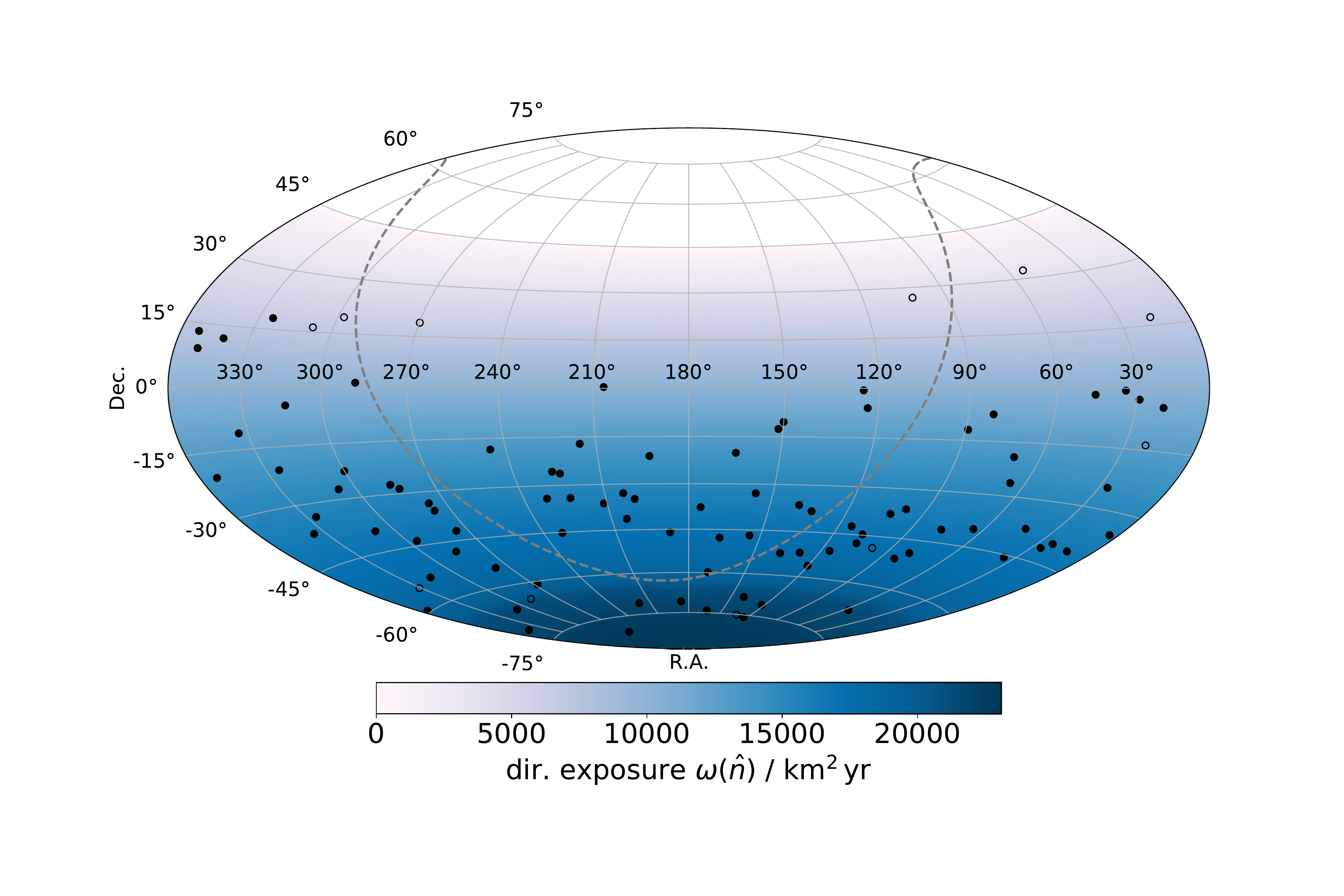}
\caption{
The positions of the arrival directions of the 100 highest-energy cosmic rays detected at the Pierre Auger Observatory shown in equatorial coordinates. The open circles show the arrival directions of the inclined events. The graded color scale shows how the exposure varies with declination for the whole data set. The white region, above $\sim$45$^{\circ}$, is not accessible from the latitude of Malargüe. The dashed line indicates the Galactic Plane.}

\label{fig:skyMap_top100.pdf}
\end{figure}

A map showing the right ascension and declinations of the 100 highest-energy events is displayed in Figure~\ref{fig:skyMap_top100.pdf}.
% created on 2022-09-19
\section*{Acknowledgments}

\begin{sloppypar}
The successful installation, commissioning, and operation of the Pierre
Auger Observatory would not have been possible without the strong
commitment and effort from the technical and administrative staff in
Malarg\"ue. We are very grateful to the following agencies and
organizations for financial support:
\end{sloppypar}

\begin{sloppypar}
Argentina -- Comisi\'on Nacional de Energ\'\i{}a At\'omica; Agencia Nacional de
Promoci\'on Cient\'\i{}fica y Tecnol\'ogica (ANPCyT); Consejo Nacional de
Investigaciones Cient\'\i{}ficas y T\'ecnicas (CONICET); Gobierno de la
Provincia de Mendoza; Municipalidad de Malarg\"ue; NDM Holdings and Valle
Las Le\~nas; in gratitude for their continuing cooperation over land
access; Australia -- the Australian Research Council; Belgium -- Fonds
de la Recherche Scientifique (FNRS); Research Foundation Flanders (FWO);
Brazil -- Conselho Nacional de Desenvolvimento Cient\'\i{}fico e Tecnol\'ogico
(CNPq); Financiadora de Estudos e Projetos (FINEP); Funda\c{c}\~ao de Amparo \`a
Pesquisa do Estado de Rio de Janeiro (FAPERJ); S\~ao Paulo Research
Foundation (FAPESP) Grants No.~2019/10151-2, No.~2010/07359-6 and
No.~1999/05404-3; Minist\'erio da Ci\^encia, Tecnologia, Inova\c{c}\~oes e
Comunica\c{c}\~oes (MCTIC); Czech Republic -- Grant No.~MSMT CR LTT18004,
LM2015038, LM2018102, CZ.02.1.01/0.0/0.0/16{\textunderscore}013/0001402,
CZ.02.1.01/0.0/0.0/18{\textunderscore}046/0016010 and
CZ.02.1.01/0.0/0.0/17{\textunderscore}049/0008422; France -- Centre de Calcul
IN2P3/CNRS; Centre National de la Recherche Scientifique (CNRS); Conseil
R\'egional Ile-de-France; D\'epartement Physique Nucl\'eaire et Corpusculaire
(PNC-IN2P3/CNRS); D\'epartement Sciences de l'Univers (SDU-INSU/CNRS);
Institut Lagrange de Paris (ILP) Grant No.~LABEX ANR-10-LABX-63 within
the Investissements d'Avenir Programme Grant No.~ANR-11-IDEX-0004-02;
Germany -- Bundesministerium f\"ur Bildung und Forschung (BMBF); Deutsche
Forschungsgemeinschaft (DFG); Finanzministerium Baden-W\"urttemberg;
Helmholtz Alliance for Astroparticle Physics (HAP);
Helmholtz-Gemeinschaft Deutscher Forschungszentren (HGF); Ministerium
f\"ur Kultur und Wissenschaft des Landes Nordrhein-Westfalen; Ministerium
f\"ur Wissenschaft, Forschung und Kunst des Landes Baden-W\"urttemberg;
Italy -- Istituto Nazionale di Fisica Nucleare (INFN); Istituto
Nazionale di Astrofisica (INAF); Ministero dell'Istruzione,
dell'Universit\'a e della Ricerca (MIUR); CETEMPS Center of Excellence;
Ministero degli Affari Esteri (MAE); M\'exico -- Consejo Nacional de
Ciencia y Tecnolog\'\i{}a (CONACYT) No.~167733; Universidad Nacional Aut\'onoma
de M\'exico (UNAM); PAPIIT DGAPA-UNAM; The Netherlands -- Ministry of
Education, Culture and Science; Netherlands Organisation for Scientific
Research (NWO); Dutch national e-infrastructure with the support of SURF
Cooperative; Poland -- Ministry of Education and Science, grant
No.~DIR/WK/2018/11; National Science Centre, Grants
No.~2016/22/M/ST9/00198, 2016/23/B/ST9/01635, and 2020/39/B/ST9/01398;
Portugal -- Portuguese national funds and FEDER funds within Programa
Operacional Factores de Competitividade through Funda\c{c}\~ao para a Ci\^encia
e a Tecnologia (COMPETE); Romania -- Ministry of Research, Innovation
and Digitization, CNCS/CCCDI UEFISCDI, grant no. PN19150201/16N/2019 and
PN1906010 within the National Nucleus Program, and projects number
TE128, PN-III-P1-1.1-TE-2021-0924/TE57/2022 and PED289, within PNCDI
III; Slovenia -- Slovenian Research Agency, grants P1-0031, P1-0385,
I0-0033, N1-0111; Spain -- Ministerio de Econom\'\i{}a, Industria y
Competitividad (FPA2017-85114-P and PID2019-104676GB-C32), Xunta de
Galicia (ED431C 2017/07), Junta de Andaluc\'\i{}a (SOMM17/6104/UGR,
P18-FR-4314) Feder Funds, RENATA Red Nacional Tem\'atica de
Astropart\'\i{}culas (FPA2015-68783-REDT) and Mar\'\i{}a de Maeztu Unit of
Excellence (MDM-2016-0692); USA -- Department of Energy, Contracts
No.~DE-AC02-07CH11359, No.~DE-FR02-04ER41300, No.~DE-FG02-99ER41107 and
No.~DE-SC0011689; National Science Foundation, Grant No.~0450696; The
Grainger Foundation; Marie Curie-IRSES/EPLANET; European Particle
Physics Latin American Network; and UNESCO.
\end{sloppypar}

\clearpage
\bibliography{top100}{}

\begin{thebibliography}{}
\expandafter\ifx\csname natexlab\endcsname\relax\def\natexlab#1{#1}\fi
\providecommand{\url}[1]{\href{#1}{#1}}
\providecommand{\dodoi}[1]{doi:~\href{http://doi.org/#1}{\nolinkurl{#1}}}
\providecommand{\doeprint}[1]{\href{http://ascl.net/#1}{\nolinkurl{http://ascl.net/#1}}}
\providecommand{\doarXiv}[1]{\href{https://arxiv.org/abs/#1}{\nolinkurl{https://arxiv.org/abs/#1}}}

\bibitem[{Aab {et~al.}(2014{\natexlab{a}})}]{PierreAuger:2014sui}
Aab, A., {et~al.} 2014{\natexlab{a}}, Phys. Rev. D, 90, 122005,
  \dodoi{10.1103/PhysRevD.90.122005}

\bibitem[{Aab {et~al.}(2014{\natexlab{b}})}]{PierreAuger:2014gko}
---. 2014{\natexlab{b}}, Phys. Rev. D, 90, 122006,
  \dodoi{10.1103/PhysRevD.90.122006}

\bibitem[{Aab {et~al.}(2014{\natexlab{c}})}]{PierreAuger:2014jss}
---. 2014{\natexlab{c}}, JCAP, 08, 019, \dodoi{10.1088/1475-7516/2014/08/019}

\bibitem[{Aab {et~al.}(2015{\natexlab{a}})}]{PierreAuger:2015eyc}
---. 2015{\natexlab{a}}, Nucl. Instrum. Meth. A, 798, 172,
  \dodoi{10.1016/j.nima.2015.06.058}

\bibitem[{Aab {et~al.}(2015{\natexlab{b}})}]{PierreAuger:2015xho}
---. 2015{\natexlab{b}}, JCAP, 08, 049, \dodoi{10.1088/1475-7516/2015/08/049}

\bibitem[{Aab {et~al.}(2016)}]{PierreAuger:2016tar}
---. 2016, Phys. Rev. D, 93, 072006, \dodoi{10.1103/PhysRevD.93.072006}

\bibitem[{Aab {et~al.}(2017{\natexlab{a}})}]{PierreAuger:2017pzq}
---. 2017{\natexlab{a}}, Science, 357, 1266, \dodoi{10.1126/science.aan4338}

\bibitem[{Aab {et~al.}(2017{\natexlab{b}})}]{PierreAuger:2017tlx}
---. 2017{\natexlab{b}}, Phys. Rev. D, 96, 122003,
  \dodoi{10.1103/PhysRevD.96.122003}

\bibitem[{Aab {et~al.}(2017{\natexlab{c}})}]{PierreAuger:2016ppv}
---. 2017{\natexlab{c}}, Astrophys. J. Lett., 837, L25,
  \dodoi{10.3847/2041-8213/aa61a5}

\bibitem[{Aab {et~al.}(2017{\natexlab{d}})}]{PierreAuger:2017vtr}
---. 2017{\natexlab{d}}, JINST, 12, P02006,
  \dodoi{10.1088/1748-0221/12/02/P02006}

\bibitem[{Aab {et~al.}(2019{\natexlab{a}})}]{PierreAuger:2019ens}
---. 2019{\natexlab{a}}, JCAP, 10, 022, \dodoi{10.1088/1475-7516/2019/10/022}

\bibitem[{Aab {et~al.}(2019{\natexlab{b}})}]{PierreAuger:2019dhr}
---. 2019{\natexlab{b}}, Phys. Rev. D, 100, 082003,
  \dodoi{10.1103/PhysRevD.100.082003}

\bibitem[{Aab {et~al.}(2020{\natexlab{a}})}]{PierreAuger:2020qqz}
---. 2020{\natexlab{a}}, Phys. Rev. D, 102, 062005,
  \dodoi{10.1103/PhysRevD.102.062005}

\bibitem[{Aab {et~al.}(2020{\natexlab{b}})}]{PierreAuger:2020yab}
---. 2020{\natexlab{b}}, JINST, 15, P10021,
  \dodoi{10.1088/1748-0221/15/10/P10021}

\bibitem[{Abraham {et~al.}(2010{\natexlab{a}})}]{PierreAuger:2010zof}
Abraham, J., {et~al.} 2010{\natexlab{a}}, Nucl. Instrum. Meth. A, 613, 29,
  \dodoi{10.1016/j.nima.2009.11.018}

\bibitem[{Abraham {et~al.}(2010{\natexlab{b}})}]{PierreAuger:2009esk}
---. 2010{\natexlab{b}}, Nucl. Instrum. Meth. A, 620, 227,
  \dodoi{10.1016/j.nima.2010.04.023}

\bibitem[{Abreu {et~al.}(2011)}]{PierreAuger:2011yxe}
Abreu, P., {et~al.} 2011, JCAP, 11, 022, \dodoi{10.1088/1475-7516/2011/11/022}

\bibitem[{Abreu {et~al.}(2012)}]{PierreAuger:2012rhm}
---. 2012, JINST, 7, P09001, \dodoi{10.1088/1748-0221/7/09/P09001}

\bibitem[{Abreu {et~al.}(2022)}]{PierreAuger:2022axr}
---. 2022, Astrophys. J., 935, 170, \dodoi{10.3847/1538-4357/ac7d4e}

\bibitem[{Alves~Batista {et~al.}(2019)Alves~Batista, Biteau, Bustamante, Dolag,
  Engel, Fang, Kampert, Kostunin, Mostafa, Murase, Oikonomou, Olinto, Panasyuk,
  Sigl, Taylor, \& Unger}]{Batista}
Alves~Batista, R., Biteau, J., Bustamante, M., {et~al.} 2019, Frontiers in
  Astronomy and Space Sciences, 6, \dodoi{10.3389/fspas.2019.00023}

\bibitem[{Andringa {et~al.}(2011)Andringa, Conceicao, \&
  Pimenta}]{Andringa:2011zz}
Andringa, S., Conceicao, R., \& Pimenta, M. 2011, Astropart. Phys., 34, 360,
  \dodoi{10.1016/j.astropartphys.2010.10.002}

\bibitem[{Ave {et~al.}(2000)Ave, Hinton, Vazquez, Watson, \& Zas}]{Ave:2000nd}
Ave, M., Hinton, J.~A., Vazquez, R.~A., Watson, A.~A., \& Zas, E. 2000, Phys.
  Rev. Lett., 85, 2244, \dodoi{10.1103/PhysRevLett.85.2244}

\bibitem[{Ave {et~al.}(2013)}]{AIRFLY:2012msg}
Ave, M., {et~al.} 2013, Astropart. Phys., 42, 90,
  \dodoi{10.1016/j.astropartphys.2012.12.006}

\bibitem[{Bonifazi(2009)}]{Bonifazi:2009ma}
Bonifazi, C. 2009, Nucl. Phys. B Proc. Suppl., 190, 20,
  \dodoi{10.1016/j.nuclphysbps.2009.03.063}

\bibitem[{Coleman {et~al.}(2022)}]{Coleman}
Coleman, A., {et~al.} 2022.
\newblock \doarXiv{2205.05845}

\bibitem[{Dawson(2020)}]{Dawson:2020bkp}
Dawson, B. 2020, PoS, ICRC2019, 231, \dodoi{10.22323/1.358.0231}

\bibitem[{Dembinski {et~al.}(2010)Dembinski, Billoir, Deligny, \&
  Hebbeker}]{Dembinski}
Dembinski, H., Billoir, P., Deligny, O., \& Hebbeker, T. 2010, Astroparticle
  Physics, 34, 128, \dodoi{https://doi.org/10.1016/j.astropartphys.2010.06.006}

\bibitem[{{Gaisser} \& {Hillas}(1977)}]{gaisser1977}
{Gaisser}, T.~K., \& {Hillas}, A.~M. 1977, in International Cosmic Ray
  Conference, Vol.~8, 353

\bibitem[{Galbraith(1958)}]{Galbraith}
Galbraith, W. 1958, Extensive Air Showers (Butterworths Publication Limited),
  185

\bibitem[{Greisen(1956)}]{Greisen1956}
Greisen, K. 1956, Progress in Cosmic Ray Physics, Vol.~3 (North-Holland
  Publishing), 1

\bibitem[{Greisen(1960)}]{Greisen1960}
---. 1960, Annual Review of Nuclear Science, 10, 63,
  \dodoi{10.1146/annurev.ns.10.120160.000431}

\bibitem[{Hillas(1977)}]{hillas1970}
Hillas, A.~M. 1977, Acta Phys. Acad. Sci. Hung., Suppl 29, 355

\bibitem[{Hillas {et~al.}(1971)Hillas, Hollows, \& Hunter}]{hillas1971}
Hillas, A.~M., Hollows, J., \& Hunter, H. 1971, in International Cosmic Ray
  Conference, Vol.~3, 1001

\bibitem[{Kamata \& Nishimura(1958)}]{Kamata}
Kamata, K., \& Nishimura, J. 1958, Progress of Theoretical Physics Supplement,
  6, 93, \dodoi{10.1143/PTPS.6.93}

\bibitem[{Linsley \& Scarsi(1962)}]{Linsley1962}
Linsley, J., \& Scarsi, L. 1962, Phys. Rev., 128, 2384,
  \dodoi{10.1103/PhysRev.128.2384}

\bibitem[{Luce {et~al.}(2021)Luce, Roth, Schmidt, \& Veberic}]{Luce:2021ode}
Luce, Q., Roth, M., Schmidt, D., \& Veberic, D. 2021, PoS, ICRC2021, 435,
  \dodoi{10.22323/1.395.0435}

\bibitem[{Mollerach \& Roulet(2018)}]{Mollerach}
Mollerach, S., \& Roulet, E. 2018, Progress in Particle and Nuclear Physics,
  98, 85, \dodoi{https://doi.org/10.1016/j.ppnp.2017.10.002}

\bibitem[{Mostaf\'{a}(2005)}]{Mostafa}
Mostaf\'{a}, M.~A. 2005, in International Cosmic Ray Conference, Vol.~7, 369.
\newblock \url{https://cds.cern.ch/record/965354}

\bibitem[{Newton {et~al.}(2007)Newton, Knapp, \& Watson}]{Newton:2006wy}
Newton, D., Knapp, J., \& Watson, A.~A. 2007, Astropart. Phys., 26, 414,
  \dodoi{10.1016/j.astropartphys.2006.08.003}

\bibitem[{Schmidt(2010)}]{Schmidt}
Schmidt, T. 2010, PhD thesis, Karlsruhe Institute of Technology

\bibitem[{Unger {et~al.}(2008)Unger, Dawson, Engel, Schussler, \&
  Ulrich}]{Unger:2008uq}
Unger, M., Dawson, B.~R., Engel, R., Schussler, F., \& Ulrich, R. 2008, Nucl.
  Instrum. Meth. A, 588, 433, \dodoi{10.1016/j.nima.2008.01.100}

\bibitem[{Valiño {et~al.}(2010)Valiño, Alvarez-Muñiz, Roth, Vazquez, \&
  Zas}]{VALINO2010304}
Valiño, I., Alvarez-Muñiz, J., Roth, M., Vazquez, R., \& Zas, E. 2010,
  Astroparticle Physics, 32, 304,
  \dodoi{https://doi.org/10.1016/j.astropartphys.2009.09.008}

\bibitem[{Yushkov(2020)}]{Yushkov:2020nhr}
Yushkov, A. 2020, PoS, ICRC2019, 482, \dodoi{10.22323/1.358.0482}

\end{thebibliography}
\bibliographystyle{aasjournal}

\begin{wrapfigure}[9]{l}{0.06\linewidth}
\vspace{-3.2ex}
\includegraphics[width=1.72\linewidth]{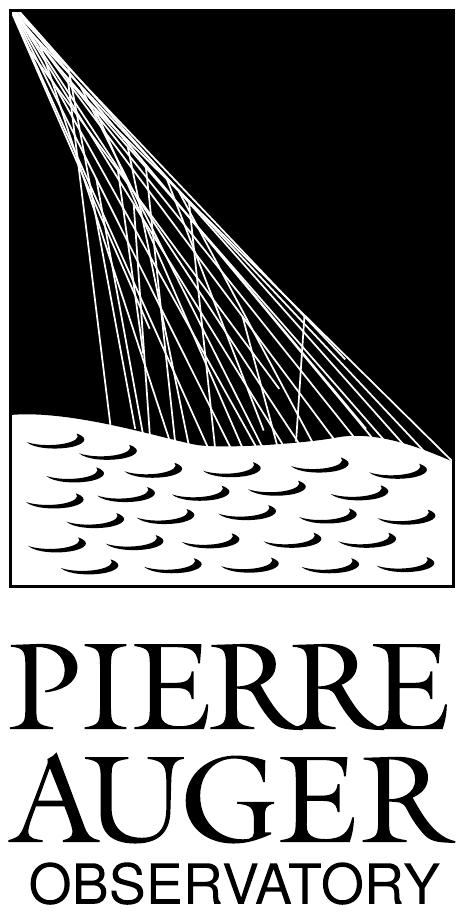}
\end{wrapfigure}
\begin{sloppypar}\noindent
% created on 2022-09-19
A.~Abdul Halim$^{13}$,
P.~Abreu$^{71}$,
M.~Aglietta$^{53,51}$,
I.~Allekotte$^{1}$,
P.~Allison$^{a}$,
K.~Almeida Cheminant$^{69}$,
A.~Almela$^{8,12}$,
J.~Alvarez-Mu\~niz$^{78}$,
J.~Ammerman Yebra$^{78}$,
G.A.~Anastasi$^{53,51}$
% ...

\end{sloppypar}
\begin{center}
\par\noindent
\textbf{The Pierre Auger Collaboration}
\end{center}

\vspace{1ex}
\begin{center}
\rule{0.1\columnwidth}{0.5pt}
\raisebox{-0.4ex}{\scriptsize$\bullet$}
\rule{0.1\columnwidth}{0.5pt}
\end{center}

\vspace{1ex}
% created on 2022-09-19
% needs \usepackage{enumitem}
\begin{enumerate}[labelsep=0.2em,align=right,labelwidth=0.7em,labelindent=0em,leftmargin=2em,noitemsep]
\item[$^{1}$] Centro At\'omico Bariloche and Instituto Balseiro (CNEA-UNCuyo-CONICET), San Carlos de Bariloche, Argentina
\item[$^{2}$] Centro de Investigaciones en L\'aseres y Aplicaciones, CITEDEF and CONICET, Villa Martelli, Argentina
\item[$^{3}$] Departamento de F\'\i{}sica and Departamento de Ciencias de la Atm\'osfera y los Oc\'eanos, FCEyN, Universidad de Buenos Aires and CONICET, Buenos Aires, Argentina
\item[$^{4}$] IFLP, Universidad Nacional de La Plata and CONICET, La Plata, Argentina
\item[$^{5}$] Instituto de Astronom\'\i{}a y F\'\i{}sica del Espacio (IAFE, CONICET-UBA), Buenos Aires, Argentina
\item[$^{6}$] Instituto de F\'\i{}sica de Rosario (IFIR) -- CONICET/U.N.R.\ and Facultad de Ciencias Bioqu\'\i{}micas y Farmac\'euticas U.N.R., Rosario, Argentina
\item[$^{7}$] Instituto de Tecnolog\'\i{}as en Detecci\'on y Astropart\'\i{}culas (CNEA, CONICET, UNSAM), and Universidad Tecnol\'ogica Nacional -- Facultad Regional Mendoza (CONICET/CNEA), Mendoza, Argentina
\item[$^{8}$] Instituto de Tecnolog\'\i{}as en Detecci\'on y Astropart\'\i{}culas (CNEA, CONICET, UNSAM), Buenos Aires, Argentina
\item[$^{9}$] International Center of Advanced Studies and Instituto de Ciencias F\'\i{}sicas, ECyT-UNSAM and CONICET, Campus Miguelete -- San Mart\'\i{}n, Buenos Aires, Argentina
\item[$^{10}$] Observatorio Pierre Auger, Malarg\"ue, Argentina
\item[$^{11}$] Observatorio Pierre Auger and Comisi\'on Nacional de Energ\'\i{}a At\'omica, Malarg\"ue, Argentina
\item[$^{12}$] Universidad Tecnol\'ogica Nacional -- Facultad Regional Buenos Aires, Buenos Aires, Argentina
\item[$^{13}$] University of Adelaide, Adelaide, S.A., Australia
\item[$^{14}$] Universit\'e Libre de Bruxelles (ULB), Brussels, Belgium
\item[$^{15}$] Vrije Universiteit Brussels, Brussels, Belgium
\item[$^{16}$] Centro Federal de Educa\c{c}\~ao Tecnol\'ogica Celso Suckow da Fonseca, Nova Friburgo, Brazil
\item[$^{17}$] Instituto Federal de Educa\c{c}\~ao, Ci\^encia e Tecnologia do Rio de Janeiro (IFRJ), Brazil
\item[$^{18}$] Universidade de S\~ao Paulo, Escola de Engenharia de Lorena, Lorena, SP, Brazil
\item[$^{19}$] Universidade de S\~ao Paulo, Instituto de F\'\i{}sica de S\~ao Carlos, S\~ao Carlos, SP, Brazil
\item[$^{20}$] Universidade de S\~ao Paulo, Instituto de F\'\i{}sica, S\~ao Paulo, SP, Brazil
\item[$^{21}$] Universidade Estadual de Campinas, IFGW, Campinas, SP, Brazil
\item[$^{22}$] Universidade Estadual de Feira de Santana, Feira de Santana, Brazil
\item[$^{23}$] Universidade Federal do ABC, Santo Andr\'e, SP, Brazil
\item[$^{24}$] Universidade Federal do Paran\'a, Setor Palotina, Palotina, Brazil
\item[$^{25}$] Universidade Federal do Rio de Janeiro, Instituto de F\'\i{}sica, Rio de Janeiro, RJ, Brazil
\item[$^{26}$] Universidade Federal do Rio de Janeiro (UFRJ), Observat\'orio do Valongo, Rio de Janeiro, RJ, Brazil
\item[$^{27}$] Universidade Federal Fluminense, EEIMVR, Volta Redonda, RJ, Brazil
\item[$^{28}$] Universidad de Medell\'\i{}n, Medell\'\i{}n, Colombia
\item[$^{29}$] Universidad Industrial de Santander, Bucaramanga, Colombia
\item[$^{30}$] Charles University, Faculty of Mathematics and Physics, Institute of Particle and Nuclear Physics, Prague, Czech Republic
\item[$^{31}$] Institute of Physics of the Czech Academy of Sciences, Prague, Czech Republic
\item[$^{32}$] Palacky University, RCPTM, Olomouc, Czech Republic
\item[$^{33}$] CNRS/IN2P3, IJCLab, Universit\'e Paris-Saclay, Orsay, France
\item[$^{34}$] Laboratoire de Physique Nucl\'eaire et de Hautes Energies (LPNHE), Sorbonne Universit\'e, Universit\'e de Paris, CNRS-IN2P3, Paris, France
\item[$^{35}$] Univ.\ Grenoble Alpes, CNRS, Grenoble Institute of Engineering Univ.\ Grenoble Alpes, LPSC-IN2P3, 38000 Grenoble, France
\item[$^{36}$] Universit\'e Paris-Saclay, CNRS/IN2P3, IJCLab, Orsay, France
\item[$^{37}$] Bergische Universit\"at Wuppertal, Department of Physics, Wuppertal, Germany
\item[$^{38}$] Karlsruhe Institute of Technology (KIT), Institute for Experimental Particle Physics, Karlsruhe, Germany
\item[$^{39}$] Karlsruhe Institute of Technology (KIT), Institut f\"ur Prozessdatenverarbeitung und Elektronik, Karlsruhe, Germany
\item[$^{40}$] Karlsruhe Institute of Technology (KIT), Institute for Astroparticle Physics, Karlsruhe, Germany
\item[$^{41}$] RWTH Aachen University, III.\ Physikalisches Institut A, Aachen, Germany
\item[$^{42}$] Universit\"at Hamburg, II.\ Institut f\"ur Theoretische Physik, Hamburg, Germany
\item[$^{43}$] Universit\"at Siegen, Department Physik -- Experimentelle Teilchenphysik, Siegen, Germany
\item[$^{44}$] Gran Sasso Science Institute, L'Aquila, Italy
\item[$^{45}$] INFN Laboratori Nazionali del Gran Sasso, Assergi (L'Aquila), Italy
\item[$^{46}$] INFN, Sezione di Catania, Catania, Italy
\item[$^{47}$] INFN, Sezione di Lecce, Lecce, Italy
\item[$^{48}$] INFN, Sezione di Milano, Milano, Italy
\item[$^{49}$] INFN, Sezione di Napoli, Napoli, Italy
\item[$^{50}$] INFN, Sezione di Roma ``Tor Vergata'', Roma, Italy
\item[$^{51}$] INFN, Sezione di Torino, Torino, Italy
\item[$^{52}$] Istituto di Astrofisica Spaziale e Fisica Cosmica di Palermo (INAF), Palermo, Italy
\item[$^{53}$] Osservatorio Astrofisico di Torino (INAF), Torino, Italy
\item[$^{54}$] Politecnico di Milano, Dipartimento di Scienze e Tecnologie Aerospaziali , Milano, Italy
\item[$^{55}$] Universit\`a del Salento, Dipartimento di Matematica e Fisica ``E.\ De Giorgi'', Lecce, Italy
\item[$^{56}$] Universit\`a dell'Aquila, Dipartimento di Scienze Fisiche e Chimiche, L'Aquila, Italy
\item[$^{57}$] Universit\`a di Catania, Dipartimento di Fisica e Astronomia ``Ettore Majorana'', Catania, Italy
\item[$^{58}$] Universit\`a di Milano, Dipartimento di Fisica, Milano, Italy
\item[$^{59}$] Universit\`a di Napoli ``Federico II'', Dipartimento di Fisica ``Ettore Pancini'', Napoli, Italy
\item[$^{60}$] Universit\`a di Palermo, Dipartimento di Fisica e Chimica ``E.\ Segr\`e'', Palermo, Italy
\item[$^{61}$] Universit\`a di Roma ``Tor Vergata'', Dipartimento di Fisica, Roma, Italy
\item[$^{62}$] Universit\`a Torino, Dipartimento di Fisica, Torino, Italy
\item[$^{63}$] Benem\'erita Universidad Aut\'onoma de Puebla, Puebla, M\'exico
\item[$^{64}$] Unidad Profesional Interdisciplinaria en Ingenier\'\i{}a y Tecnolog\'\i{}as Avanzadas del Instituto Polit\'ecnico Nacional (UPIITA-IPN), M\'exico, D.F., M\'exico
\item[$^{65}$] Universidad Aut\'onoma de Chiapas, Tuxtla Guti\'errez, Chiapas, M\'exico
\item[$^{66}$] Universidad Michoacana de San Nicol\'as de Hidalgo, Morelia, Michoac\'an, M\'exico
\item[$^{67}$] Universidad Nacional Aut\'onoma de M\'exico, M\'exico, D.F., M\'exico
\item[$^{68}$] Universidad Nacional de San Agustin de Arequipa, Facultad de Ciencias Naturales y Formales, Arequipa, Peru
\item[$^{69}$] Institute of Nuclear Physics PAN, Krakow, Poland
\item[$^{70}$] University of \L{}\'od\'z, Faculty of High-Energy Astrophysics,\L{}\'od\'z, Poland
\item[$^{71}$] Laborat\'orio de Instrumenta\c{c}\~ao e F\'\i{}sica Experimental de Part\'\i{}culas -- LIP and Instituto Superior T\'ecnico -- IST, Universidade de Lisboa -- UL, Lisboa, Portugal
\item[$^{72}$] ``Horia Hulubei'' National Institute for Physics and Nuclear Engineering, Bucharest-Magurele, Romania
\item[$^{73}$] Institute of Space Science, Bucharest-Magurele, Romania
\item[$^{74}$] University Politehnica of Bucharest, Bucharest, Romania
\item[$^{75}$] Center for Astrophysics and Cosmology (CAC), University of Nova Gorica, Nova Gorica, Slovenia
\item[$^{76}$] Experimental Particle Physics Department, J.\ Stefan Institute, Ljubljana, Slovenia
\item[$^{77}$] Universidad de Granada and C.A.F.P.E., Granada, Spain
\item[$^{78}$] Instituto Galego de F\'\i{}sica de Altas Enerx\'\i{}as (IGFAE), Universidade de Santiago de Compostela, Santiago de Compostela, Spain
\item[$^{79}$] IMAPP, Radboud University Nijmegen, Nijmegen, The Netherlands
\item[$^{80}$] Nationaal Instituut voor Kernfysica en Hoge Energie Fysica (NIKHEF), Science Park, Amsterdam, The Netherlands
\item[$^{81}$] Stichting Astronomisch Onderzoek in Nederland (ASTRON), Dwingeloo, The Netherlands
\item[$^{82}$] Universiteit van Amsterdam, Faculty of Science, Amsterdam, The Netherlands
\item[$^{83}$] Case Western Reserve University, Cleveland, OH, USA
\item[$^{84}$] Colorado School of Mines, Golden, CO, USA
\item[$^{85}$] Department of Physics and Astronomy, Lehman College, City University of New York, Bronx, NY, USA
\item[$^{86}$] Michigan Technological University, Houghton, MI, USA
\item[$^{87}$] New York University, New York, NY, USA
\item[$^{88}$] University of Chicago, Enrico Fermi Institute, Chicago, IL, USA
\item[$^{89}$] University of Delaware, Department of Physics and Astronomy, Bartol Research Institute, Newark, DE, USA
\item[$^{90}$] University of Wisconsin-Madison, Department of Physics and WIPAC, Madison, WI, USA
\item[] -----
\item[$^{a}$] Ohio State University, Columbus, OH, USA
\item[$^{b}$] Louisiana State University, Baton Rouge, LA, USA
\item[$^{c}$] also at University of Bucharest, Physics Department, Bucharest, Romania
\item[$^{d}$] School of Physics and Astronomy, University of Leeds, Leeds, United Kingdom
\item[$^{e}$] now at Deutsches Elektronen-Synchrotron DESY, Zeuthen, Germany
\item[$^{f}$] Fermi National Accelerator Laboratory, Fermilab, Batavia, IL, USA
\item[$^{g}$] now at Graduate School of Science, Osaka Metropolitan University, Osaka, Japan
\item[$^{h}$] Max-Planck-Institut f\"ur Radioastronomie, Bonn, Germany
\item[$^{i}$] Colorado State University, Fort Collins, CO, USA
\item[$^{j}$] Pennsylvania State University, University Park, PA, USA
\item[$^{k}$] Los Alamos National Laboratory, Los Alamos, NM, USA
\item[$^{l}$] Laboratoire AstroParticule et Cosmologie (APC), Université Paris 7, CNRS-IN2P3, Paris, France
\end{enumerate}

\end{document}